\documentclass[adp,fleqn]{w-art}
\usepackage{times}
\usepackage{w-thm}
\usepackage[]{graphicx}
\usepackage{cite}
\begin{document}
\DOIsuffix{theDOIsuffix}
\Volume{12}
\Issue{1}
\Copyrightissue{01}
\Month{01}
\Year{2003}
\pagespan{1}{}
\Receiveddate{18 March 2004}
\Accepteddate{26 March 2004 by U.\ Eckern}
 % up to three, separated by commas
\title[]{The vanadium Magn\'eli phases V${\bf_n}$O${\bf_{2n-1}}$}
\author[Udo Schwingenschl\"ogl]{Udo Schwingenschl\"ogl\footnote{Corresponding
     author \quad E-mail: {\sf Udo.Schwingenschloegl@Physik.Uni-Augsburg.de}, Fax: +49\,821\,598\,3262\\
     Online colour figures at: www.ann-phys.org}} 
\address[]{Institut f\"ur Physik, Universit\"at Augsburg, 86135 Augsburg, Germany}
\author[Volker Eyert]{Volker Eyert}
\renewcommand{\leftmark}{U.\ Schwingenschl\"ogl, V.\ Eyert: The vanadium Magn\'eli phases V$_n$O$_{2n-1}$}

\begin{abstract}
To compare the metal-insulator transitions (MITs) of VO$_2$ and V$_2$O$_3$
we analyze the relations between the structural and electronic properties of the vanadium Magn\'eli phases. These
materials set up the homologous series V$_n$O$_{2n-1}$ $(3\leq n\leq9)$ and have crystal structures comprising
typical dioxide-like and sesquioxide-like regions. As the MITs of the vanadium Magn\'eli
phases are accompanied by structural transformations, we are able to discuss the effects of characteristic
changes in the local atomic environments. The systematic investigation of the transport properties is based on a
new and unifying description of the crystal structures of the Magn\'eli phases including VO$_2$ and V$_2$O$_3$. 
Our results lead to a comprehensive understanding of the MITs in the Magn\'eli class and
shed new light on the role of particular electronic states for the MIT of V$_2$O$_3$. 
\end{abstract}
\maketitle
\tableofcontents
\section{Introduction}
The vanadium oxides \cite{goodenough71,brueckner83,imada98}
comprise compounds with different formal vanadium valency stages,
which reach from two in VO, three in V$_2$O$_3$, and four in VO$_2$ to five in V$_2$O$_5$.
In addition to these configurations mixed valent compounds can be synthesized.
Amongst the latter the so-called Magn\'eli phases, defined by the general stoichiometric formula
\begin{equation}\label{eq23}
{\rm V}_n{\rm O}_{2n-1}={\rm V}_2{\rm O}_3+(n-2){\rm VO}_2\,\;\;\;{\rm where}\;\;\;3\leq n\leq 9\,,
\end{equation}
are of special interest since they give rise to a homologous series of compounds with closely related
crystal structures. This type of homologous series has been reported for the first time by Magn\'eli 
for the molybdenum oxides \cite{magneli48}. Nowadays additional Magn\'eli series are known for
the vanadium, titanium, niobium, and tungsten oxides. The first structural characterization of
vanadium Magn\'eli compounds by means of x-ray investigations traces back to Andersson and Jahnberg
\cite{andersson63} in 1963.
As equation (\ref{eq23}) indicates, the Magn\'eli phases
take an intermediary position between V$_2$O$_3$ and VO$_2$, thus between the
valency stages three and four. In addition to the chemical relationsship, the crystal structures of the
Magn\'eli phases actually consist of rutile and corundum-type blocks and consequently show structural
affinity to both the dioxide and the sesquioxide. In section \ref{sec4} we study the structural
relations between these materials in detail -- discovering the potential of gradually
transferring the crystal structure of VO$_2$ into that of V$_2$O$_3$ by making use of the Magn\'eli phases.
This observation allows for a comprehensive understanding of all these compounds.

As a function of temperature each Magn\'eli phase undergoes a metal-insulator transition (MIT), except for
V$_7$O$_{13}$, which is metallic at all temperatures. The MITs are of first order and accompanied by distinct
structural transformations.
Table \ref{tab1} summarizes transition temperatures, inferred from electrical resistivity measurements by Kachi et
al.\ \cite{kachi73}, and formal V $3d$ charges,
varying from two in V$_2$O$_3$ to one in VO$_2$. In an ionic picture a metal atom
contributes three $3d$ and two $4s$ electrons, whereas an oxygen atom accepts two $2p$ electrons.
As the Magn\'eli phases are characterized by vanadium atoms in mixed valent
states, the electronic properties were expected to be influenced by charge ordering. Although this is
interesting from both the experimental and theoretical point of view, only few
studies are reported in the literature.

\begin{SCfigure}[10][t!]
\includegraphics[width=8.6cm,clip]{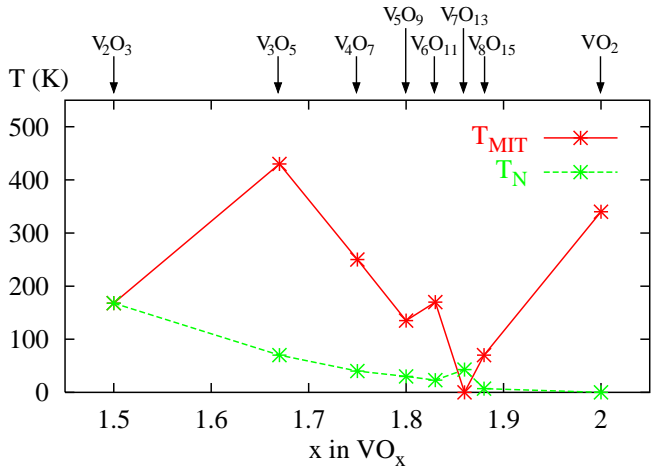}
\caption{Comparison of the MIT temperatures and the magnetic ordering temperatures
of the Magn\'eli phases. Besides V$_7$O$_{13}$ all members of the series undergo an MIT.
Except for VO$_2$ each compound develops an antiferromagnetic order at low temperatures. In the 
case of V$_2$O$_3$ the MIT and the magnetic ordering coincide at a temperature of 168\,K.}
\label{pic36}\end{SCfigure}

Figure \ref{pic36} illustrates the behaviour of the MIT temperatures listed in table \ref{tab1} as a function of the
vanadium-oxygen ratio. We recognize a broad minimum centered at the ratio corresponding to V$_7$O$_{13}$.
\begin{table}[t!]\begin{center}
\begin{tabular}{|c|c|c|c|}\hline
Compound V$_n$O$_{2n-1}$ & Parameter n & Formal V $3d$ charge & MIT temperature\\\hline
V$_2$O$_3$ & 2 & 2 & 168\,K\\
V$_3$O$_5$ & 3 & 5/3 $\approx 1.67$ & 430\,K\\
V$_4$O$_7$ & 4 & 6/4 $\approx 1.50$ & 250\,K\\
V$_5$O$_9$ & 5 & 7/5 $\approx 1.40$ & 135\,K\\
V$_6$O$_{11}$ & 6 & 8/6 $\approx 1.33$ & 170\,K\\
V$_7$O$_{13}$ & 7 & 9/7 $\approx 1.29$ & metallic\\
V$_8$O$_{15}$ & 8 & 10/8 $\approx 1.25$ & 70\,K\\
V$_9$O$_{17}$ & 9 & 11/9 $\approx 1.22$ & ---\\
VO$_2$ & $\infty$ & 1 & 340\,K\\\hline
\end{tabular}
\caption{Formal V $3d$ charges as well as transition temperatures in the series V${_n}$O$_{2n-1}$.}
\label{tab1}\end{center}\end{table}
Interestingly, the MITs of the vanadium Magn\'eli phases are coupled to an anomaly in the
magnetic susceptibility, which closely resembles the characteristics of a transition from a paramagnetic
to an antiferromagnetic state \cite{kachi73}. Nonetheless, the anomaly is not accompanied by long range
magnetic order. Yet, all Magn\'eli phases enter an
antiferromagnetic groundstate at sufficiently low temperatures; N\'eel temperatures are given in
figure \ref{pic36}. In some sense the anomaly may be
interpreted as a precursor effect. There is a maximum of the transition temperature for
V$_2$O$_3$ and a gradual decrease on approaching VO$_2$,
where antiferromagnetic ordering is no longer observed. V$_7$O$_{13}$ is the only deviation
from this general trend. Although a continuous reduction of V$_2$O$_3$-like regions in the crystal structure
is a possible explanation for decreasing transition temperatures,
this kind of mechanism has not been confirmed. While MIT and magnetic ordering are coupled
for V$_2$O$_3$, they appear at different temperatures and thus
are likely to occur independently for other Magn\'eli phases. As we are interested in the MITs,
the subsequent electronic structure calculations do not take into
account the low temperature antiferromagnetism. Actually, it would not be possible to include
the magnetic order since the detailed magnetic structures of the ordered phases are not known.

Specific heat data for the Magn\'eli phases by Khattak et
al.\ \cite{khattak78} revealed values larger than expected from pure lattice
contributions. Due to the smallness of the magnetic entropy increase at the N\'eel temperatures
the authors argued in favour of magnetic correlations above the phase transitions. To
explain their findings they proposed a model based on linear antiferromagnetic chains. However,
low temperature investigations of the magnetic susceptibility by Nagata et al.\ \cite{nagata79} ruled out this model.
The authors suggested interpreting insulating V$_8$O$_{15}$ in terms of a charge-density wave,
as Gossard et al.\ \cite{gossard74} speculated in the case of V$_4$O$_7$. They
attributed small magnetic moments at the metal sites of insulating V$_4$O$_7$/V$_6$O$_{11}$ to singlet pairing.

Single crystal x-ray data by Marezio et al.\ \cite{marezio72} indicated chains of V$^{3+}$ and V$^{4+}$ ions
in the insulating phase of V$_4$O$_7$, whereas the vanadium valences are disordered
in the metallic phase. This kind of evidence for charge localization arises from the
comparison of experimental V-O distances in V$_4$O$_7$ with estimated values referring to an ionic picture.
VO$_6$ units with large/small V-O bond lengths are interpreted in terms of V$^{3+}$/V$^{4+}$ configurations,
respectively. However, due to the covalent portion of the interatomic bonding the calculated
charges might be poor. Specific heat data by Griffing et al.\ \cite{griffing82,griffing85}
implied that metallic V$_4$O$_7$ and V$_7$O$_{13}$ are characterized by a coexistence of itinerant and
localized V $3d$ electrons. Nuclear magnetic resonance experiments confirmed this result
and additionally found two inequivalent metal sites, but were in conflict with pure
$S=1/2$ and $S=1$ spin states, corresponding to $d^1$ and $d^2$ configurations \cite{gossard74, gossard74a}.
In the insulating phase the Magn\'eli compounds show indications of increased charge localization and singlet pairing,
except for V$_3$O$_5$. The charge localization may split the subbands associated with
the inequivalent metal sites and hence may pave the way for a Mott--Hubbard transition.

While the MIT temperature of VO$_2$ increases on the application of pressure, V$_2$O$_3$ reveals the
opposite trend. The Magn\'eli phases reflect the behavior of the sesquioxide as their transition temperatures decrease
with growing hydrostatic pressure \cite{brueckner83}. Moreover, Canfield et al.\ \cite{canfield90a}
reported on unifying aspects emerging in the Magn\'eli series on the application of pressure. They observed
for both V$_8$O$_{15}$ and (V$_{0.98}$Cr$_{0.02}$)$_8$O$_{15}$ a transition from a paramagnetic metal (PM phase) to
an antiferromagnetic metal (AFM phase) and further to a paramagnetic insulator (PI phase).
Consequently, the PM-AFM transition present for V$_7$O$_{13}$ at atmospheric pressure should not
be regarded anomalous for the Magn\'eli series. While the PM-AFM transition is found in V$_8$O$_{15}$ only
above $9$\,kbar, it is always present for the chromium doped system.
The magnetic transition exhibits the same pressure dependence in all three cases
($-0.75$\,K/kbar). Since the MIT temperature decreases at a larger rate, the AFM modification is stabilized
for sufficiently high pressures. Although in other Magn\'eli phases the PM-AFM transition is not observed up to
$20$\,kbar, the authors assume different pressure dependences of the MIT and the antiferromagnetic ordering as a general
feature, common to each Magn\'eli phase. This again indicates that the transitions evolve independently.

\section{Crystal structure and electronic properties of ${\bf VO_2}$ and ${\bf V_2O_3}$ }
\begin{SCfigure}[10][t!]
\includegraphics[width=8.0cm,clip]{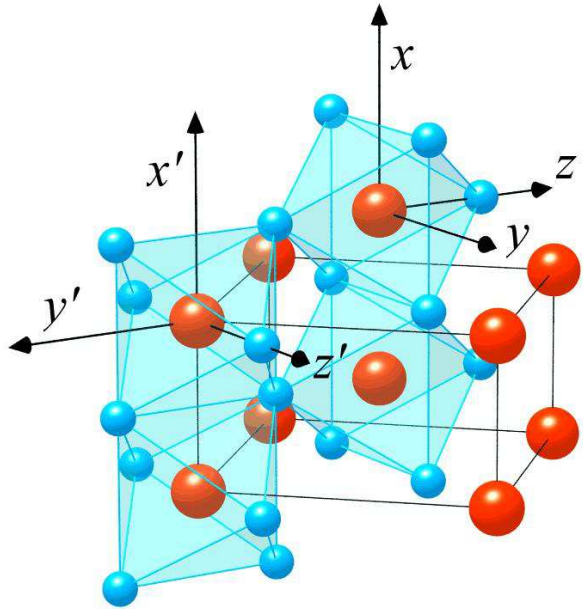}
\caption{Crystal structure of metallic VO${_2}$. The inserted cuboid determines the
unit cell of the rutile structure with the $c$-axis oriented upwards.
Large and small spheres represent vanadium and oxygen atoms, respectively. The vanadium
atoms form infinite chains along the rutile $c$-axis. Furthermore, each vanadium
site is surrounded by an oxygen octahedron. The latter are oriented uniformly
in each vanadium chain but rotated by 90\,$^{\circ}$ with respect
to the rutile $c$-axis among neighbouring chains. Hence they establish local coordinate systems.}
\label{pic1}\end{SCfigure}

Above 340\,K ${\rm VO_2}$ crystallizes in the rutile crystal structure based on a
simple tetragonal lattice with space group $P4_2/mnm$ $(D_{4h}^{14})$, as shown in figure \ref{pic1}.
Using single crystal experiments McWhan et al.\ \cite{mcwhan74} determined the tetragonal lattice
constants $a_R=4.5546$\,\AA\ and $c_R=2.8514$\,\AA\ and the positional parameters.
With respect to the vanadium sites the ${\rm VO_2}$ structure is based on a body-centered tetragonal
lattice, where each metal site is surrounded by an oxygen octahedron. Octahedra at the center and
the corners of a rutile unit cell are rotated by 90$^{\circ}$ around the rutile $c$-axis. This
reduces the lattice symmetry from body-centered tetragonal to simple tetragonal and causes
the unit cell to contain two formular units. Equally adjusted octahedra form chains along the rutile
$c$-axis. Within the chains neighbouring octahedra share edges, whereas corner and center chains are
connected via corners. Next to the filled oxygen octahedra the ${\rm VO_2}$ structure
comprises the same amount of empty octahedra, which likewise form chains along the rutile $c$-axis.
Hence ${\rm VO_2}$ is based on a regular three dimensional network of oxygen octahedra partially
filled with vanadium atoms. Each octahedron has orthorhombic symmetry, but deviations from tetragonal and
even cubic symmetry are small. One observes two different V-O distances, where the apical distance appears
twice within each octahedron and the equatorial distance is found four times \cite{sorantin92}.

We define local coordinate systems centered at the metal sites. Because of the different orientations of the
oxygen octahedra two reference systems are required.
In both cases the local $z$-axis is oriented along the apical axis of the (local) ocahedron, which is either the (110)
or the (1$\bar{1}$0) direction. Compared to the traditional alignment of the local $x$ and
$y$-axis parallel to the metal-ligand bonds, these axes are rotated by 45$^{\circ}$ around the
local $z$-axis. Thus they are parallel and perpendicular to the rutile $c$-axis, respectively. This definition
of a local rotated reference frame is useful not only for rutile VO$_2$ but also for
related structures. Referring to the original rutile coordinates we introduce the pseudorutile axes $a_{\rm prut}$,
$b_{\rm prut}$ and $c_{\rm prut}$, which allows us to discuss the crystal structures of the different Magn\'eli
phases on a common basis.

\begin{SCfigure}[10][t!]
\includegraphics[width=8.8cm,clip]{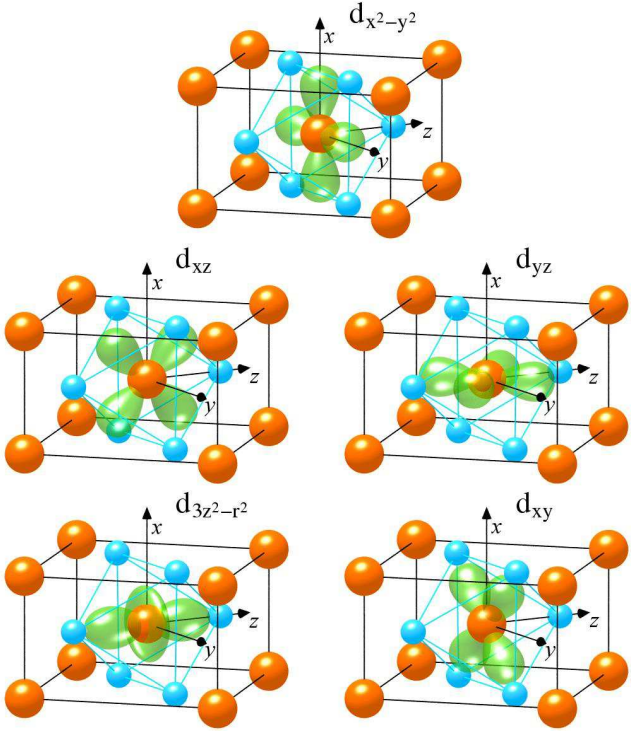}
\caption{Angular parts of the transition metal $3d$ states with respect to the local coordinate system of the central metal
atom in the displayed rutile unit cell. Note that the rutile $c$-axis is oriented upwards.
The ${t_{2g}}$ manifold, resulting from the crystal field splitting of the metal states, comprises the
${d_{x^2-y^2}}$, ${d_{xz}}$ and ${d_{yz}}$ orbitals, whereas the ${e_g}$ manifold
comprises the ${d_{3z^2-r^2}}$ and ${d_{xy}}$ states. While the ${d_{x^2-y^2}}$
lobes point at edges of the local octahedron, the ${d_{xz}}$ and
${d_{yz}}$ orbitals are directed towards faces. Both ${e_g}$ orbitals point at oxygen atoms.}
\label{pic4}\end{SCfigure}

Figure \ref{pic4} displays the angular parts of the metal $d$ orbitals relative to the reference frame of the
central metal atom. The cubic part of the crystal field splitting results in a separation of the $d$
orbitals in threefold degenerate $t_{2g}$ ($d_{x^2-y^2}$, $d_{xz}$, $d_{yz}$) and twofold degenerate $e_g$
($d_{3z^2-r^2}$, $d_{xy}$) states. While both $e_g$ orbitals point to oxygen atoms,
the $d_{x^2-y^2}$ orbital points along the rutile $c$-axis (local $x$-axis) and the local $y$-axis.
Moreover, $d_{xz}$/$d_{yz}$ orbitals are directed to faces of the local oxygen octahedron. Hence
$\sigma$ and $\pi$-type V-V overlap along the rutile $c$-axis is mediated by $d_{x^2-y^2}$ and $d_{xz}$ states,
respectively. Due to the 45$^{\circ}$ rotation of our coordinate system the $d_{x^2-y^2}$
and $d_{xy}$ orbitals are interchanged compared to the standard notation. The lobes of the $d_{yz}$ orbital
point perpendicular to the rutile $c$-axis. Because of the ratio $c_R$/$a_R=0.6260$ of the crystal
axes they exhibit reduced $\sigma$-type overlap, compared to the $d_{x^2-y^2}$
states. While both $d_{x^2-y^2}$ and $d_{yz}$ orbitals connect metal atoms
separated by tetragonal lattice vectors, coupling between metal sites at the center and the corners
of the unit cell is mediated by $d_{xz}$ orbitals, which exhibit remarkable
overlap with the $d_{x^2-y^2}$ states belonging to the vanadium atoms of neighbouring octahedral chains.

The first order phase transition of stoichiometric VO$_2$ at 340\,K is accompanied by structural distortions leading to the
monoclinic M$_1$ phase, which is
based on a simple monoclinic lattice with space group $P2_1/c$ $(C_{2h}^{5})$ \cite{morin59}.
Despite very similar gross features, distinct differences distinguish rutile and monoclinic VO$_2$.
First, a characteristic metal-metal pairing along the $c_{\rm prut}$-axis is present
in the monoclinic structure, which modifies the V-V distances (2.851\,\AA\ in the rutile case)
and gives rise to alternating values of 2.619\,\AA\ and 3.164\,\AA. Second,
zigzag-type displacements of the metal sites evolve along the diagonal
of the rutile basal plane (local $z$-axis). The shift direction alternates along $a_{\rm prut}$/$c_{\rm prut}$
but not along $b_{\rm prut}$. Due to the zigzag-type distortions two different apical V-O
bond lengths of 1.77\,\AA\ and 2.01\,\AA\ are observed. In addition, the V-V pairing results in two short
(1.86\,\AA\ and 1.89\,\AA) and two long (2.03\,\AA\ and 2.06\,\AA) equatorial distances in each oxygen
octahedron. Third, a lattice strain is present in the monoclinic configuration.

Uniaxial stress along the rutile (110)
axis or doping with some percent of chromium/aluminium yields two additional insulating
phases: the monoclinic M$_2$ and triclinic T phase \cite{marezio72a,pouget75,pouget76}.
In the M$_1$ phase V-V pairing and zigzag-type lateral displacements of the metal atoms affect each
vanadium chain. In contrast, in the M$_2$ phase half of the chains dimerize and the other half show zigzag-type deviations.
The T phase is intermediate since the dimerized M$_2$ chains start to tilt and the zigzag chains evolve a dimerization
until eventually the M$_1$ structure is reached.

An instability of the rutile crystal structure is found not only for VO$_2$ but also
applies to other transition metal oxides, as for instance MoO$_2$ and NbO$_2$. For the latter compounds
the relations between structural distortions and electronic structure changes have been
studied in \cite{eyert00,eyert02}. As VO$_2$, they are characterized by metal-metal pairing
and lateral zigzag-type displacements in the low temperature phase. Despite all the
similarities a striking difference regarding the in-plane distortions of the metal atoms and the modifications
of the surrounding oxygen octahedra distinguishes monoclinic VO$_2$ and MoO$_2$. Only for the
latter material the octahedra follow the in-plane metal shifts, whereas for VO$_2$ the vanadium atoms move
relative to their oxygen octahedra. This relative shift affects the bonding between V $3d$ and O $2p$ states as it
influences the orbital overlap. Thus an antiferroelectric distortion of the VO$_6$ octahedra
is inherent in the monoclinic (M$_1$) phase of VO$_2$. From the different behaviour of
VO$_2$ and MoO$_2$ it was concluded that the antiferroelectric distortion cannot explain the
tendency of the transition metal oxides to form distorted variants of the ideal rutile structure. For the same
reason the mechanism must be excluded as a possible driving force for MITs connected to
the destabilization of the rutile crystal structure \cite{eyert02a}.

Theoretical approaches towards understanding the MIT and the simultaneous structural transformations
range from Peierls \cite{peierls55,gruener88} to Mott--Hubbard
\cite{mott68,gebhard97} schemes. Possible driving forces are lattice
instabilities, electron-phonon interaction, and electronic correlations.
The importance of the lattice for stabilization of the different phases of VO$_2$
and the symmetry change at the transition point to strong electron-phonon
coupling \cite{mcwhan74}. This line of reasoning is supported by a comparison of rutile and
monoclinic VO$_2$ via ultrasonic microscopy, which reveals a strong elastic anisotropy in the
metallic but almost no anisotropy in the insulating phase \cite{maurer99}. The
electronic structure of metallic VO$_2$ has been probed by optical measurements revealing the lowest empty V $3d$
levels 2.5\,eV above the top edge of the O $2p$ bands \cite{verleur68}. Ultraviolet
and x-ray photoelectron spectroscopy show a 8.5\,eV wide occupied band directly below the
Fermi energy \cite{goering97}. The valence band splits up into low and high binding
regions with widths of 1.5\,eV and 6\,eV, respectively. While the low binding contributions are
attributed to the V $3d$ states, the broader part of the valence band mainly traces back to O $2p$ states.
According to oxygen K-edge x-ray absorption spectroscopy unoccupied V $3d$ bands extend from the
Fermi energy to 1.7\,eV and from 2.2\,eV to 5.2\,eV \cite{mueller97}. For monoclinic VO$_2$
photoelectron spectroscopy shows a sharpening accompanied by an energetical downshift of the occupied
V $3d$ bands. Shin et al.\ reported a band gap of $0.7$\,eV for the insulating phase \cite{shin90}.

Goodenough proposed an energy band scheme for both metallic and insulating VO$_2$
\cite{goodenough71}. Starting with electrostatic considerations he placed the
O $2p$ states well below the V $3d$ states. Octahedral crystal field splitting leads to lower V $3d$ $t_{2g}$ and higher
V $3d$ $e_g$ levels, where the former are located in the vicinity of the Fermi level and split into $d_{\parallel}$
(oriented parallel to the rutile $c$-axis) and $\pi^{*}$ states. In the monoclinic case
metal-metal pairing causes the $d_{\parallel}$ band
to split into filled bonding and empty antibonding states. Due to antiferroelectric
zigzag-type displacements of the metal atoms, the $\pi^{*}$ bands shift to higher energies.
Shin et al.\ found the $d_{\parallel}$ band splitting to amount to 2.5\,eV and the
$\pi^{*}$ bands to rise by 0.5\,eV \cite{shin90}. In contrast to the Goodenough scheme Zylbersztejn and Mott proposed an MIT
mechanism based on strong electron-electron correlations \cite{zylbersztejn75}. According to these authors
especially the one-dimensional $d_{\parallel}$ band is affected by correlations rather than by electron-lattice interaction.
In the metallic phase correlations are efficiently screened by the $\pi^{*}$ electrons, but screening
is reduced below the phase transition as the $\pi^{*}$ bands are subject to an energetical upshift.
Thus the narrow $d_{\parallel}$ bands become
susceptible to strong Coulomb repulsion and undergo a Mott-transition -- opening a correlation gap.
Recently, the MIT of VO$_2$ was explained in terms of a Peierls instability of the
one-dimensional $d_{\parallel}$ band, embedded in a background of the remaining V $3d$ $t_{2g}$
states \cite{eyert02a,eyert98}. Molecular dynamics calculations by Wentzcovitch et al.\ likewise point to
predominant influence of the lattice degrees of freedom \cite{wentzcovitch94}.

Stoichiometric V$_2$O$_3$ at ambient pressure undergoes an MIT at 168\,K accompanied by
a transformation from the high temperature corundum into the low temperature monoclinic structure, where additionally
an antiferromagnetic order appears (AFI phase).
Systematic studies have provided the temperature, pressure, and composition dependence of the MIT in the
(V$_{1-x}$Cr$_x$)$_2$O$_3$ system \cite{mcwhan71,mcwhan73}. Doping with chromium or aluminum results in a
paramagnetic insulating configuration (PI phase), which retains the lattice symmetry of the paramagnetic metal
(PM phase). Since the metal-insulator boundary terminates at a critical point, a gradual PM-PI crossover
appears when decreasing the temperature from well above the 500\,K region. For a
chromium concentration $0.005\leq x_{\rm Cr}\leq0.018$ three phase transitions are encountered: PM-PI, PI-PM,
and PM-AFI, where the last two are of first order \cite{kuwamoto80}. For $x_{\rm Cr}<0.005$ the reentrant MIT
disappears. In contrast, on doping with titanium the PM-AFI transition temperature decreases and for $x_{\rm Ti}>0.05$
only a metallic phase is found.
Hydrostatic pressure has almost the same effect as doping with titanium.

\begin{figure}[t!]\begin{center}
\includegraphics[width=10.0cm,clip]{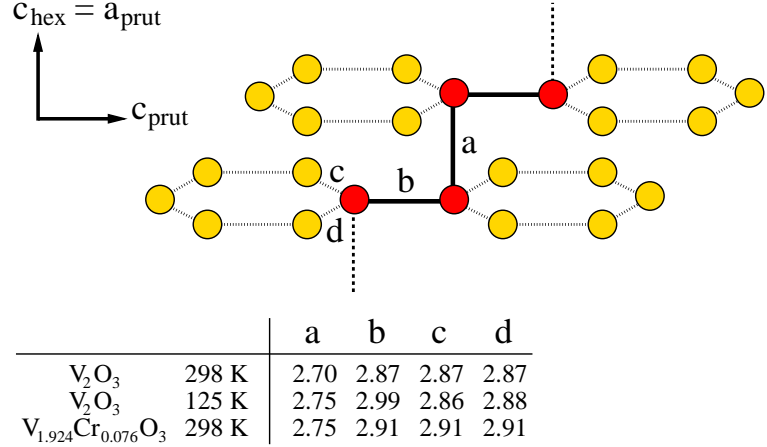}
\caption{Crystal structure of V$_2$O$_3$. For simplicity the schematic view does not include the
oxygen sublattice, which is an octahedral network qualitatively equal to the case of
VO$_2$. The vanadium atoms once more occupy octahedral sites but now form
hexagonal structures perpendicular to the c${_{\rm hex}}$-axis. Hence vanadium
pairs arise both along ${a_{\rm prut}}$=${c_{\rm hex}}$ and ${c_{\rm prut}}$. The table records experimental
V-V distances (in \AA) for the paramagnetic metallic, the antiferromagnetic insulating, and the paramagnetic
insulating phase of V$_2$O$_3$ \cite{dernier70,dernier70a}.}
\label{pic10}\end{center}\end{figure}

Corundum V$_2$O$_3$ is based on a trigonal lattice with space group $R\bar{3}c$
($D^6_{3d}$). Oftentimes a non-primitive hexagonal unit cell comprising six formula units
is used, whereas the primitive trigonal unit cell contains two formula units. The metal atoms are octahedrally
coordinated by six oxygen atoms, where the oxygen network is qualitatively equivalent to the arrangement in VO$_2$.
We identify the $c_{\rm hex}$-axis of the hexagonal unit cell with the $a_{\rm prut}$-axis.
Two thirds of the oxygen octahedra are filled with vanadium, but no continuous
chains of face-sharing VO$_6$ octahedra arise. Instead, two filled octahedra are followed by an empty one.
Neighbouring chains are shifted along c$_{\rm hex}$, giving rise to hexagonal vanadium structures, see figure \ref{pic10}.
This forms the basis of describing the corundum structure in terms of a hexagonal cell.
Because of the trigonal symmetry each metal atom exhibits three identical in-plane V-V distances of $2.87$\,\AA.
Metal-metal electrostatic interaction along c$_{\rm hex}$ (through octahedral faces) causes V-V anti-dimerization
reflected by a bond length of $2.70$\,\AA, which is considerably larger than the ideal
octahedral separation of $2.33$\,\AA. Hence the hexagonal vanadium structures deviate from being planar.
There are three V-O bond lengths of $1.97$\,\AA\ and three other of $2.05$\,\AA\ in each VO$_6$ unit.

At 168\,K stoichiometric V$_2$O$_3$ distorts into a monoclinic structure with space group $I2/a$ $(C_{2h}^6)$,
characterized by increasing nearest neighbour V-V distances \cite{dernier70a}. Bond lengths
across shared octahedral faces along a$_{\rm prut}$ increase from $2.70$\,\AA\ to $2.75$\,\AA.
Moreover, the threefold degeneracy of the in-plane V-V distance is lifted, giving rise
to a strongly elongated ($2.99$\,\AA) bond along $c_{\rm prut}$-axis.
The symmetry breaking at the PM-AFI transition results in two crystallographically inequivalent oxygen sites, where
each oxygen octahedron comprises four sites of the first and two of the second kind. Although the V-O distance
varies from $1.95$\,\AA\ to $2.11$\,\AA, the average length is similar to the PM phase.
In contrast to the symmetry breaking at the PM-AFI transition, the corundum structure is
conserved at the PM-PI transition \cite{dernier70}. Nevertheless, V-V bond lengths across
octahedral faces/edges increase to $2.75$\,\AA/$2.91$\,\AA. A comparison
of PI (aluminum doped) and AFI V$_2$O$_3$, as based on hard and soft x-ray absorption spectroscopy, revealed the same local symmetry for the
metal sites and hence (on a local scale) identical distortions \cite{pfalzer02}.

V$_2$O$_3$ is generally regarded as canonical Mott--Hubbard system
\cite{mott68, gebhard97}. A description in terms of the one-band Hubbard model is based
on a level scheme by Castellani et al.\ \cite{castellani78}:
Crystal field splitting of the V $3d$ states due to the octahedral coordination yields lower $t_{2g}$ and
higher $e_g^{\sigma}$ states. Because of the trigonal lattice
symmetry the former split up into $a_{1g}$ and $e_g^{\pi}$ contributions. The non-degenerate
$a_{1g}$ orbital points along $c_{\rm hex}$ and gives rise to a one-dimensional band due to covalent bonding.
This electronic level scheme was confirmed by band structure calculations \cite{mattheiss94}. Castellani et al.\ assumed
the bonding $a_{1g}$ orbital as fully occupied, whereas the antibonding states shift energetically above
the $e_g^{\pi}$ levels. This leaves one electron per vanadium site in the
twofold degenerate $e_g^{\pi}$ states, making the system susceptible to degeneracy lifting distortions.
One $e_g^{\pi}$ electron (total spin $S=1/2$) suggests using the half filled one-band Hubbard model
as the simplest model describing V$_2$O$_3$ \cite{rozenberg95}. 

Experimentally, the PM-AFI phase transition was analyzed by photoelectron spectroscopy. As reported by Shin et
al.\ \cite{shin90}, the spectra show O $2p$ states in the energy range from $-10$\,eV to
$-4$\,eV, whereas V $3d$ states occur within $3$\,eV below the Fermi level. Except for slight
modifications of the bandwidth no drastic change is found in the V $3d$ band structure below the MIT.
However, the DOS at the Fermi energy is rather small even in the metallic phase. A high resolution
photoemission study by Shin et al.\ \cite{shin95} revealed a band gap of $0.2$\,eV for the
insulating phase. The PM-PI transition of (chromium doped)
V$_2$O$_3$ was investigated by Smith and Henrich \cite{smith94} using photoelectron spectroscopy. In the insulating phase
at room temperature they found a low emission intensity at the Fermi level, which increases after cooling into the
metallic state. High temperature PI spectra reveal a larger V $3d$ bandwidth than low
temperature AFI spectra, which may be due to thermal broadening or
absence of magnetic order. According to oxygen K-edge x-ray absorption spectroscopy unoccupied V $3d$ bands
extend from $0$\,eV (Fermi energy) to $6$\,eV with maxima at about $1$\,eV and $3$\,eV \cite{mueller97a}.

The model of Castellani et al.\ has been called into question by the results of polarized x-ray absorption spectroscopy,
which indicate a vanadium $S=1$ spin state \cite{park00}. Furthermore, the first excited states
miss a pure $e_g^{\pi}$ symmetry but include $a_{1g}$ contributions, which
requires an explanation beyond the pure one-band Hubbard model or beyond
models projecting out the $a_{1g}$ states by means of simple dimerization. LDA+U calculations succeeded
in explaining the electronic and magnetic properties of the AFI phase, in particular the peculiar
antiferromagnetic order \cite{ezhov99}. While every vanadium spin is parallel to the adjacent spin along
$c_{\rm hex}$, there are two antiparallel spins and one parallel spin at the nearest in-plane vanadium sites.
Enforcing spin degeneracy in the calculation, influence of the crystal structure on the electronic
properties was shown to be rather small. Introducing the antiferromagnetic order yields an optical band gap of
$0.6$\,eV. Moreover, LDA+U findings point at a spin $S=1$ model without orbital degeneracy. In contrast,
by means of model calculations Mila et al.\ \cite{mila00} proposed a $S=1$ spin state and orbital
degeneracy in the AFI phase. Starting with the assumption of strong covalent bonding in the vanadium
pairs along $c_{\rm hex}$ the authors supposed the intersite $a_{1g}$ hopping matrix elements to
dominate. However, an analysis of the hopping processes in V$_2$O$_3$ revealed hopping integrals
between second, third, and fourth nearest vanadium neighbours being equally important for the shape of the $a_{1g}$
band \cite{elfimov03}. Evidence for orbital ordering in the AFI phase is taken from magnetic neutron
scattering \cite{bao97}.

LDA calculations display only minor response of
the electronic structure to crystal parameter changes. But a slight narrowing of the characteristic $a_{1g}$
bands in the insulating phase formed the basis for describing the PM-PI transition by a combination
of LDA calculations and the dynamical mean field theory (DMFT), demonstrating the influence of electronic correlations
in the insulating phase \cite{held01}. Confirming predictions of the LDA+DMFT approach, a prominent
quasiparticle peak was observed in the photoemission spectrum of metallic V$_2$O$_3$ \cite{mo03}. However,
despite many studies the electronic properties and MITs of V$_2$O$_3$ are still controversially discussed.
The answer of the electronic states
to the structural modifications at the MIT of VO$_2$ paved the way for interpreting the transition as an embedded
Peierls instability. A similar procedure seems to fail for the sesquioxide since here the structural
modifications hardly influence the LDA results.

\section{Crystal structures of the Magn\'eli phases: a unified representation}
\label{sec4}
To analyze the electronic properties of the vanadium Magn\'eli phases,
we first have to understand in detail the crystal structures --
a prerequisite for relating electronic changes at the MITs of the Magn\'eli phases to
structural transformations occuring simultaneously. We elaborate a new and unifying point of
view of the atomic arrangements underlying all members of the series. A distinct advantage of this representation
is its applicability to the crystal structures of vanadium dioxide and sesquioxide as well. Thus
a comprehensive understanding of the whole Magn\'eli series V$_n$O$_{2n-1}$ including its
end members VO$_2$ ($n\to\infty$) and V$_2$O$_3$ ($n=2$) is eventually achieved. The detailed modifications of the
various crystal structures at the MITs are discussed in subsequent sections.

\begin{figure}[t!]\begin{center}
\includegraphics[width=10.0cm,clip]{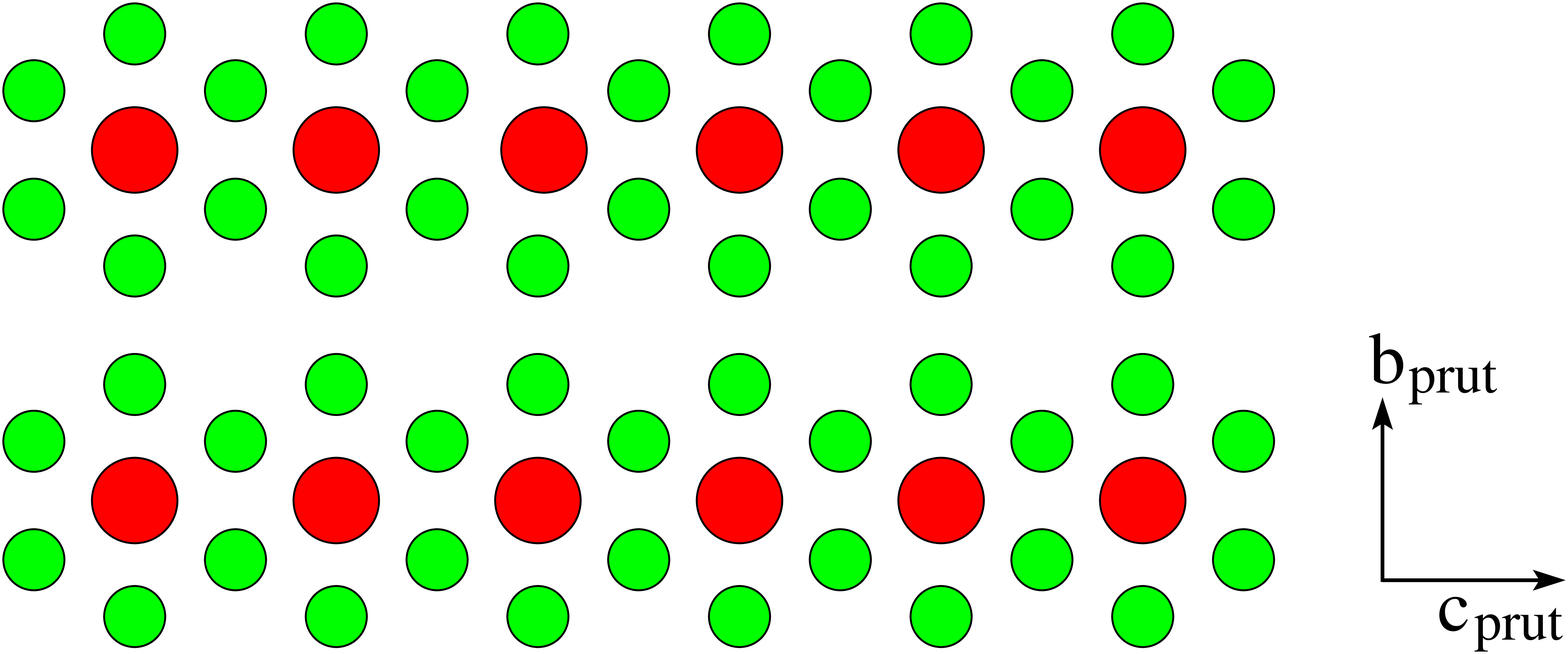}\\[0.5cm]
\includegraphics[width=10.0cm,clip]{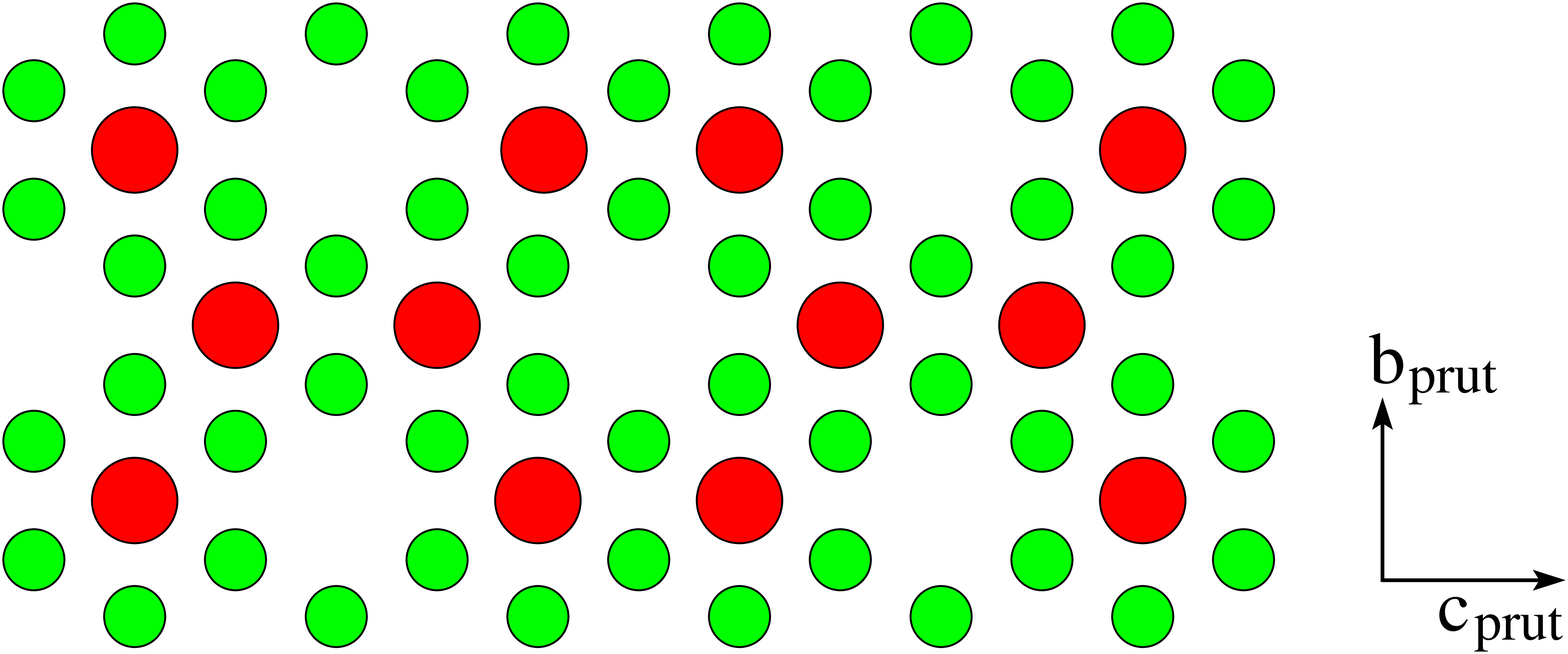}
\caption{Projection parallel to $a_{\rm prut}$ of a O-V-O sandwich-like slab cut out of the crystal structures
of VO${_2}$ (top) and V${_2}$O${_3}$ (bottom). Large and small spheres
represent vanadium and oxygen atoms, respectively. Due to the projection the oxygen octahedra form a regular
hexagonal network. For VO${_2}$ infinite chains of vanadium atoms run along the
pseudorutile $c_{\rm prut}$-axis, where the oxygen octahedra are connected by common edges. The
projection of an adjacent O-V-O slab results in a similar configuration -- but
empty and filled octahedra are exchanged. Thus the metal atoms in VO${_2}$
lack nearest neighbours in the a${_{\rm prut}}$-direction. For V${_2}$O${_3}$
finite chains of two vanadium atoms run along the $c_{\rm prut}$-axis. The projection of  an adjacent
O-V-O slab reveals a shift by one octahedral site along $c_{\rm prut}$. As a consequence,
the vanadium atoms exhibit exactly one nearest vanadium neighbour in the $a_{\rm prut}$-direction.}
\label{pic7}\end{center}\end{figure}

As mentioned before and as is obvious from the stoichiometric relation (\ref{eq23}), the
crystal structures of the Magn\'eli phases are usually interpreted as rutile-type slabs separated by shear
planes with a corundum-type atomic arrangement \cite{andersson63}. The rutile-type slabs extend infinitely in two
dimensions and have a characteristic finite width corresponding to $n$ VO$_6$ octahedra in the case of the
compound V$_n$O$_{2n-1}$. Since octahedra at a slab-surface share faces with octahedra from a neighbouring slab,
the atomic arrangement at the boundary is closely related to the corundum structure.
The rutile crystal structure is therefore disturbed and adjacent slabs are mutually out of phase, thus giving
rise to the denotation shear plane.
Precise investigation of the changes of the local atomic coordination on passing through the Magn\'eli series is
rather complicated using the ordinary representation of the crystal structures. Thus it is advantageous to
describe the structures in a different manner. For that purpose we start with the oxygen sublattice,
which turns out to be most similar for all the compounds. As demonstrated for VO$_2$ and V$_2$O$_3$
the sublattice can be understood as a regular space filling network of neighbouring oxygen octahedra. The
octahedra are mutually connected via edges along $c_{\rm prut}$ and via faces along
$a_{\rm prut}$/$b_{\rm prut}$. In order to illustrate this geometrical arrangement
schematical projections (parallel to $a_{\rm prut}$) of sandwich-like O-V-O slabs cut out of the crystal structures of the
dioxide and the sesquioxide are shown in figure \ref{pic7}.
A sandwich-like slab consists of a vanadium layer confined by oxygen layers at both ends. 
Due to the projection the oxygen octahedra appear as
hexagonal structures and the octahedral network becomes a two-dimensional hexagonal network.

Both the infinite vanadium chains of VO$_2$ (along $c_{\rm prut}$) and the hexagonal
\begin{figure}[t!]\begin{center}
\includegraphics[width=10.0cm,clip]{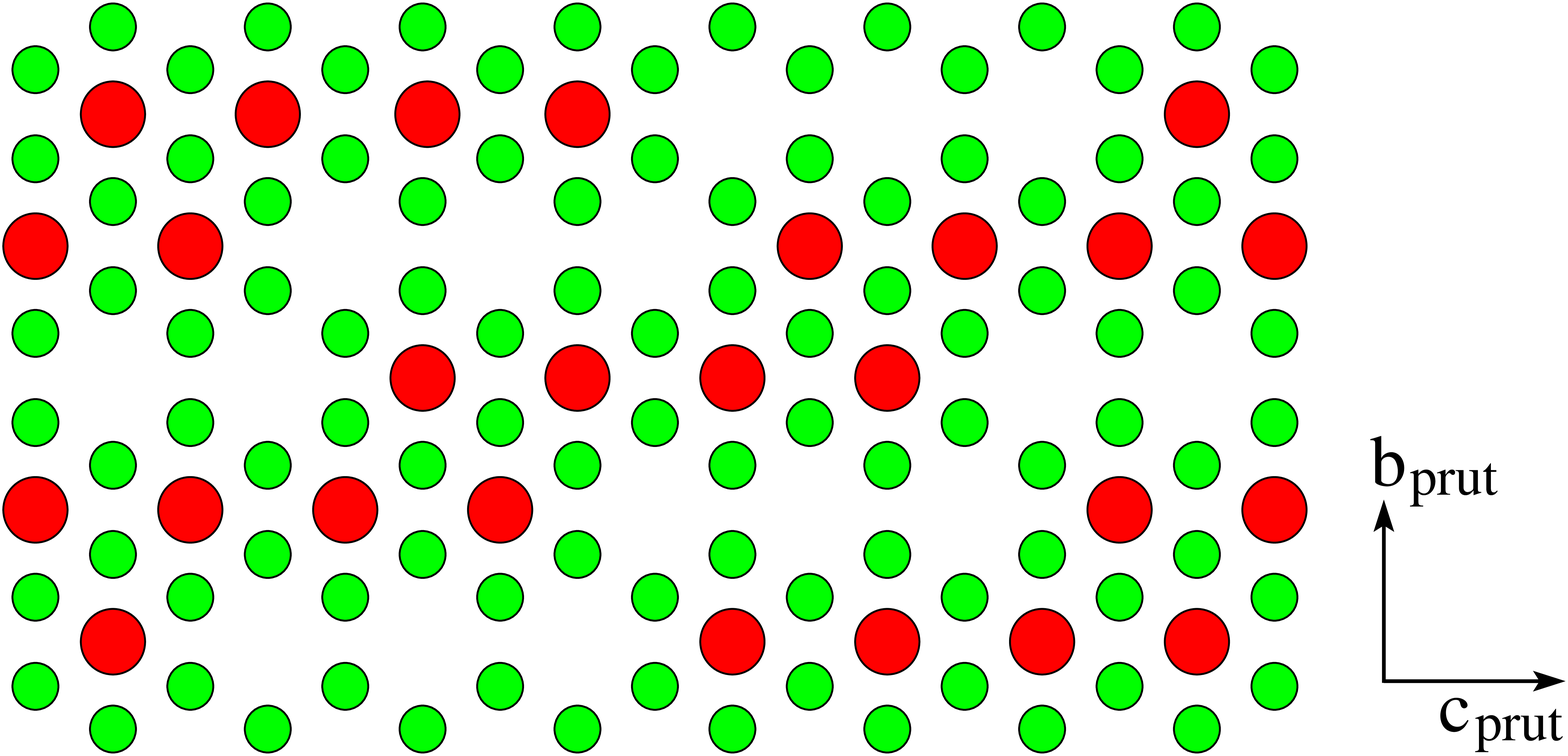}\\[0.5cm]
\includegraphics[width=10.0cm,clip]{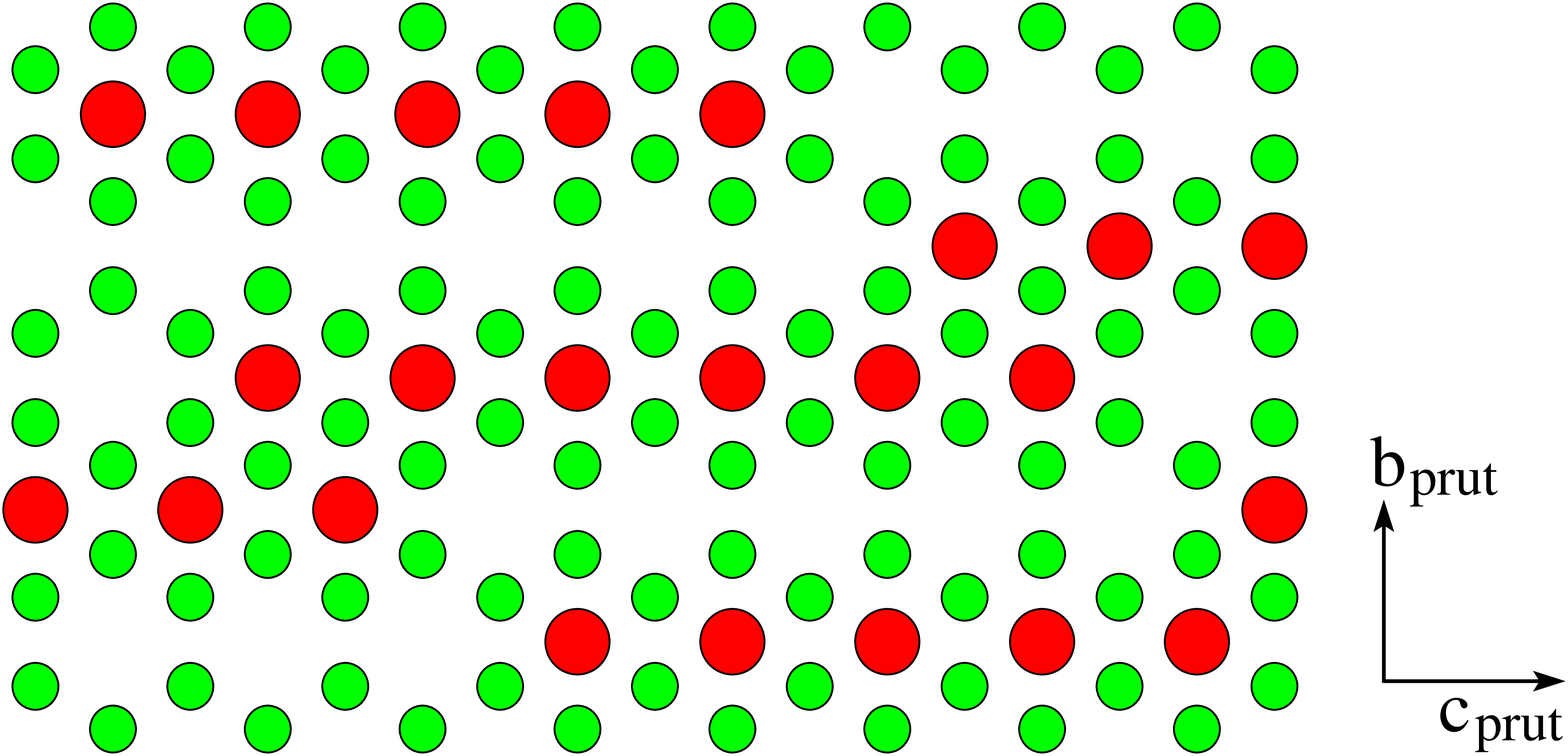}
\caption{Projection along ${a_{\rm prut}}$ of a O-V-O sandwich-like slab from the crystal structures
of V$_4$O$_7$ (top) and V$_6$O$_{11}$ (bottom). Large and small spheres represent
vanadium and oxygen atoms, respectively. As in the cases of vanadium dioxide and sesquioxide the oxygen
octahedra form a regular hexagonal network and the octahedral sites are partially occupied by metal atoms,
giving rise to vanadium chains along $c_{\rm prut}$. The chains consist of $n=4$ or $n=6$
atoms, separated by $n-1=3$ or $n-1=5$ empty octahedra. Along $b_{\rm prut}$ the chains overlap by
roughly half the intrachain V-V distance. The projection of an adjacent O-V-O slab results in a similar
configuration -- but the vanadium sublattice is shifted along $c_{\rm prut}$ by $n-1=3$ or $n-1=5$
octahedral sites. As a consequence, chain end atoms exhibit one nearest vanadium neighbour in the
$a_{\rm prut}$-direction, whereas center atoms reveal none.}
\label{pic9}\end{center}\end{figure}
arrangement of the vanadium atoms in V$_2$O$_3$ are clearly visible in figure \ref{pic7}. Actually, the basic
structural features of the dioxide
and the sesquioxide are captured completely in this graphical representation. Adjacent slabs
yield the same projection as shown except for a possible translation perpendicular to $a_{\rm prut}$.
To be more specific, in adjacent VO$_2$ slabs filled and empty oxygen octahedra, i.e.\ filled and empty
oxygen hexagons, are interchanged. In neighbouring V$_2$O$_3$ slabs
the vanadium atoms are shifted by one hexagon along the $c_{\rm prut}$-direction. To sum up, provided that
the oxygen networks of VO$_2$ and V$_2$O$_3$ are exactly the same, the only difference between the
compounds arises from a different vanadium sublattice. The configuration of the latter is
optimally described in terms of the presented slab-type projection. In particular, the infinite vanadium chains
belonging to the dioxide are contrasted with finite chains of length 2 in the case of the sesquioxide. Such a
description of the V$_2$O$_3$ structure in terms of 2-chains along $c_{\rm prut}$ instead of hexagonal
arrangements is preferable for a systematic discussion. There are small
differences between the oxygen sublattices of VO$_2$ and V$_2$O$_3$. However, they do not affect the gross
features of the oxygen arrangement, but still are
essential for a detailed comparison of the materials. As indicated by our representation in terms of
O-V-O slabs we find alternating vanadium and oxygen layers along $a_{\rm prut}$ in both compounds. While the
oxygen layers of the sesquioxide are almost flat, we are confronted with a distinct buckling in the
case of the dioxide, which is caused by an elongated apical V-O distance.
The buckling of the oxygen layers is associated with the fact that octahedral faces oriented
perpendicular to $a_{\rm prut}$ in V$_2$O$_3$ adopt a tilt in the VO$_2$ case.
Nonetheless, the characteristic feature distinguishing vanadium dioxide and sesquioxide certainly is their
different pattern of filling the oxygen network with vanadium atoms.

\begin{figure}[t!]\begin{center}
\includegraphics[width=10.0cm,clip]{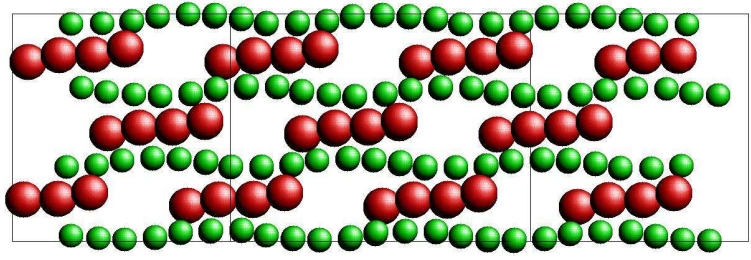}
\caption{Vanadium (large) and oxygen (small) layers in the case of the V$_4$O$_7$ crystal structure, projection
perpendicular to ${a_{\rm prut}}$. A distinct buckling of the oxygen layers is clearly visible.}
\label{pic70}\end{center}\end{figure}

Having developed a common description for the crystal structures of the end members of the
Magn\'eli series it is easy to extend this point of view to the remaining compounds. For that
purpose the structures of V$_4$O$_7$ and V$_6$O$_{11}$ are depicted in figure \ref{pic9} analogous
with the representations of VO$_2$ and V$_2$O$_3$ in figure \ref{pic7}. Except for a slightly
different buckling of the oxygen layers, which is visible in figure \ref{pic70} for V$_4$O$_7$,
the aforementioned regular three-dimensional oxygen network forms the basis of all Magn\'eli phases. Hence the hexagonal
arrangement of the oxygen atoms in the projection of the O-V-O sandwich-like slabs does not change.
In contrast, each particular material is characterized by the way of filling the oxygen network with vanadium atoms.
Obviously, there are chains of vanadium atoms running along
the $c_{\rm prut}$-axis, which comprise four and six atoms in the case of V$_4$O$_7$ and V$_6$O$_{11}$,
respectively. In general, there are vanadium chains of length $n$ in V$_n$O$_{2n-1}$.
Looking back to the dioxide ($n=\infty$) and the sesquioxide ($n=2$) this rule coincides with our finding of
infinite chains and 2-chains, respectively. In a particular compound all metal chains 
reveal the same length, defined by the parameter $n$; $n-1$ empty oxygen octahedra
separate them in the $c_{\rm prut}$-direction. Vanadium chains neighbouring along $b_{\rm prut}$ overlap
by roughly half the in-chain V-V distance, thus giving rise to the characteristical chain end arrangement
depicted in figure \ref{pic9}. This chain end configuration is conserved in the whole Magn\'eli series and
only the length of the chain center changes. For VO$_2$ there are actually no chain ends and the
2-chains of V$_2$O$_3$ contain no center atom.

\begin{figure}[t!]\begin{center}
\includegraphics[width=10.0cm,clip]{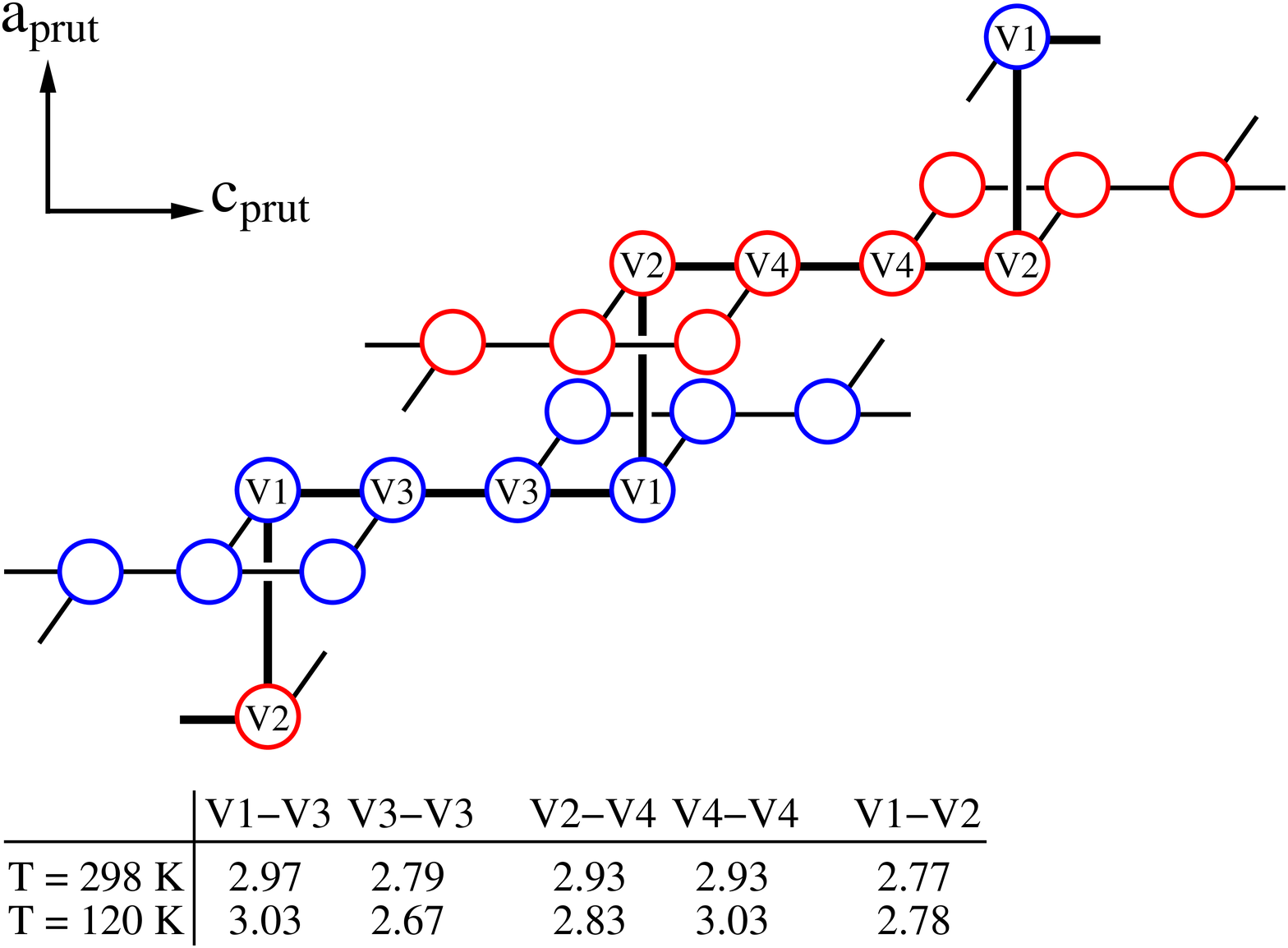}
\caption{Crystal structure of V$_4$O$_7$. For simplicity this schematic view does not include the
oxygen sublattice. The metal atoms form chains of length $n=4$ along $c_{\rm prut}$. There are 4
inequivalent sites in the chains V1-V3-V3-V1 and V2-V4-V4-V2. Layers
perpendicular to $a_{\rm prut}$ either comprise chains of the first or of the second kind.
The table gives measured V-V distances (in \AA) for both the metallic (298\,K) and insulating
(120\,K) phase of V$_4$O$_7$ \cite{hodeau78}.}
\label{pic20}\end{center}\end{figure}

\begin{figure}[t!]\begin{center}
\includegraphics[width=10.0cm,clip]{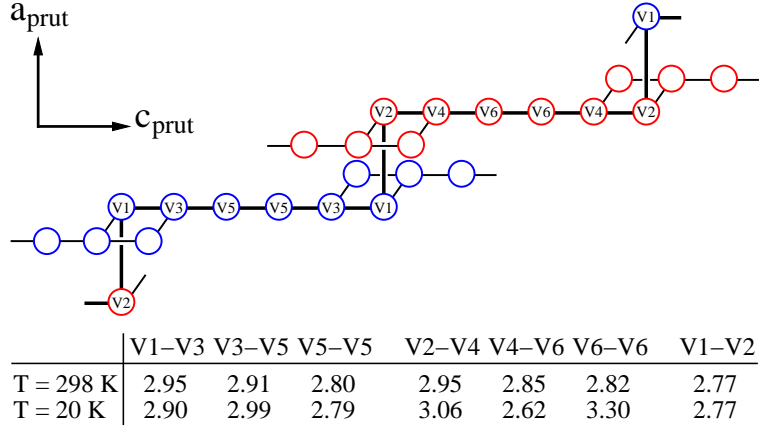}
\caption{Crystal structure of V$_6$O$_{11}$. For simplicity this schematic view does not include the
oxygen sublattice. The metal atoms form chains of length $n=6$ along $c_{\rm prut}$. There are 6
inequivalent sites in the chains V1-V3-V5-V5-V3-V1 and V2-V4-V6-V6-V4-V2. Layers
perpendicular to $a_{\rm prut}$ either comprise chains of the first or of the second kind.
The table gives measured V-V distances (in \AA) for both the metallic (298\,K) and insulating
(20\,K) phase of V$_6$O$_{11}$ \cite{canfield90}.}
\label{pic19}\end{center}\end{figure}

The atomic arrangement represented by the projections
of the O-V-O slabs is unaltered for slabs neighbouring in the $a_{\rm prut}$-direction.
Only the vanadium sublattice is shifted along the $c_{\rm prut}$-axis by an amount of $n-1$ octahedral sites,
i.e.\ $n-1$ hexagons in the sandwich-like projection. Thus the last vanadium atom of a metal chain
is seated on top of the first atom of a chain in the adjacent slab, see figures \ref{pic20} and \ref{pic19}.
Hence the chain end atoms exhibit exactly one nearest vanadium neighbour along the $a_{\rm prut}$-axis, whereas each
chain center atom is surrounded by two empty sites in this direction. All the above considerations apply
to each Magn\'eli compound. Chain center sites hence mirror the
local atomic coordination known from VO$_2$ with nearest vanadium neighbours only in the chain direction.
Instead, the coordination of the chain end atoms resembles the crystal structure of V$_2$O$_3$ with
one nearest vanadium neighbour along $a_{\rm prut}$ and $c_{\rm prut}$. Due to the varying length of the
vanadium chains in a qualitatively unchanged oxygen sublattice, the different Magn\'eli phases have
different stoichiometric V:O ratios reaching from 1:2 in VO$_2$ to 2:3 in V$_2$O$_3$. As the
crystal structures of the Magn\'eli phases are connected to each other simply by removing or
inserting chain center atoms, the Magn\'eli series allows us to transfer a dioxide-type atomic arrangement
step by step into a sesquioxide-type arrangement. Using these systematics we gain
insight into the interplay between VO$_2$ and V$_2$O$_3$-like regions and
their effects on the MIT. Taking into account the mixed valent metal states, the Magn\'eli series transfers 
the electronic configuration of the dioxide ($d^1$) into that of the sesquioxide ($d^2$).

In the following sections two particular Magn\'eli phases are discussed in detail: V$_4$O$_7$
and V$_6$O$_{11}$. To define the notation of the different metal sites figure \ref{pic20} gives a schematic view
of the V$_4$O$_7$ crystal structure where the regular network of oxygen octahedra has not been included for
simplicity. We identify vanadium 4-chains parallel to the $c_{\rm prut}$-direction. Due
to mutual interconnection of the chains along $a_{\rm prut}$ a kind of stair-like vanadium arrangement is observable.
Stair endings resemble the corundum atomic coordination (figure \ref{pic10}), whereas intermediate plateaus are due to
the rutile regions of the crystal. The longer
the central part of the vanadium chains, the larger the separation of the stair endings and the dioxide-like character
of the crystal. The illustration of V$_6$O$_{11}$ in figure \ref{pic19}
is different from figure \ref{pic20} due to longer vanadium chains.
We find four crystallographically inequivalent vanadium sites in the case of V$_4$O$_7$ and
six sites for V$_6$O$_{11}$. In both structures these sites set up two types
of metal chains: for V$_4$O$_7$ we find the series V1-V3-V3-V1 and V2-V4-V4-V2, whereas for V$_6$O$_{11}$
the series V1-V3-V5-V5-V3-V1 and V2-V4-V6-V6-V4-V2 arise. Vanadium layers
comprising either chains of the first or of the second kind alternate along $a_{\rm prut}$.

The V-V distances given as insets in the above figures address the dominating
structural changes accompanying the MIT of both V$_4$O$_7$ and V$_6$O$_{11}$
\cite{horiuchi72,marezio73,hodeau78,canfield90}. At the phase transition of the former compound the dimerization
in the 1-3 chains becomes stronger and an additional dimerization evolves in the 2-4 chains. Thereby the dimerization
patterns are reversed because the longer V-V distance appears at the chain ends and the chain center,
respectively. The V1-V2 bond length is almost constant. For V$_6$O$_{11}$ changes in the 1-3-5 chains are
small although a little dimerization is present at low temperatures. In contrast, the
2-4-6 chains evolve a pronounced dimerization at the transition. Again the variation of the V1-V2 bond
length is by far the smallest observed. A profound analysis of the structural changes
induced by the MITs is given in the subsequent sections where we relate LDA results
to local structural properties. However, signatures of the VO$_2$-like dimerization
are reflected by both V$_4$O$_7$ and V$_6$O$_{11}$.

As far as the question of choosing the unit cells of the Magn\'eli phases is concerned, several definitions have been used in the literature.
Andersson and Jahnberg \cite{andersson63} defined primitive translations in terms of the rutile lattice
$({\bf a}_A,{\bf b}_A,{\bf c}_A)=M({\bf a}_R,{\bf b}_R,{\bf c}_R)$, where they distinguished the cases
$n$ odd and $n$ even
\begin{equation}\label{eq25a}
M_{\rm odd}=
\left(\begin{array}{ccc} -1 & 0 & 1\\ 1 & 1 & 1\\ \frac{n-3}{2}&\frac{2-n}{2}&\frac{6-n}{2}\\ \end{array}\right)\,,\;\;\;
M_{\rm even}=
\left(\begin{array}{ccc} -1 & 0 & 1\\ 1 & 1 & 1\\ n-3 & 2-n & 6-n\\ \end{array}\right)\,.
\end{equation} 
However, this proposal is not useful for a comparison within the Magn\'eli series since for every parameter
value $n$ the vanadium chains extend in different directions of the unit cell, and due to the two
families of cell parameters. Le~Page and Strobel \cite{lepage82} hence proposed an alternative
set of primitive translations with the $c$-axis parallel to the vanadium chains
\begin{equation}\label{eq25}
\left(\begin{array}{c} {\bf a}_L\\ {\bf b}_L\\ {\bf c}_L\end{array}\right)=
\left(\begin{array}{ccc} -1 & 0 & 1\\ 1 & 1 & 1\\ 0 & 0 & 2n-1\\ \end{array}\right)
\left(\begin{array}{c} {\bf a}_R\\{\bf b}_R\\ {\bf c}_R\end{array}\right)\,.
\end{equation} 
While the unit cell of Andersson and Jahnberg is A-centered, the cell of Le~Page and
Strobel is I-centered. Finally, Horiuchi et al.\ \cite{horiuchi76} introduced a primitive unit cell with
translation vectors
\begin{equation}\label{eq31}
\left(\begin{array}{c} {\bf a}_H\\ {\bf b}_H\\ {\bf c}_H\end{array}\right)=
\left(\begin{array}{ccc} -1 & 0 & 1\\ 1 & 1 & 1\\ 0 & n-\frac{1}{2} & n-\frac{1}{2}\\ \end{array}\right)
\left(\begin{array}{c} {\bf a}_R\\{\bf b}_R\\ {\bf c}_R\end{array}\right)\,.
\end{equation}
Because this unit cell has half the size than the previous cells we adopt it for our calculation. 
Application of relation (\ref{eq31}) based on the parent rutile lattice is feasible whenever
the crystal symmetry is triclinic with space group $P\bar{1}$ ($C_i^1$) \cite{hahn89}, which is the case
for the Magn\'eli compounds with parameters $n=4,...,9$. In contrast, V$_3$O$_5$ crystallizes in a body-centered and
simple monoclinic lattice with space group $I2/c$ $(C_{2h}^6)$ and $P2/c$ $(C_{2h}^4)$ above and below the
MIT, respectively \cite{asbrink80,hong82}.
\section{Structural and calculational details}
The first vanadium Magn\'eli compound we discuss at length is ${\bf V_6O_{11}}$, see section
\ref{sec1}. In doing so we can benefit from a suitable application of the knowledge we already acquired by our
former considerations of VO$_2$. Consequently, it is convenient to choose V$_6$O$_{11}$ as the starting point
of a comprehensive analysis of the whole Magn\'eli class. Similar to the MIT of vanadium
dioxide and the PM-AFI transition of vanadium sesquioxide, the MIT of V$_6$O$_{11}$ at 170\,K is
accompanied by distinct structural transformation. According to an x-ray study by Canfield \cite{canfield90}, V$_6$O$_{11}$
crystallizes in a triclinic lattice with space group $P\bar{1}$ ($C_i^1$). The vanadium and oxygen
atoms are located at the Wyckoff positions (2i): $\pm (x,y,z)$. For the triclinic
lattice parameters the author reported $a_L=5.449$\,\AA, $b_L=7.010$\,\AA, $c_L=31.437$\,\AA,
$\alpha_L=67.15^{\circ}$, $\beta_L=57.45^{\circ}$, and $\gamma_L=108.90^{\circ}$ at $298$\,K.
Below the phase transition (at $20$\,K) he found the parameters $a_L=5.495$\,\AA, $b_L=6.944$\,\AA,
$c_L=31.484$\,\AA, $\alpha_L=67.40^{\circ}$, $\beta_L=57.13^{\circ}$, and $\gamma_L=108.61^{\circ}$. These
values are related to the unit cell of Le~Page and Strobel as denoted in equation (\ref{eq25}),
where $n=6$. The primitive translations of Horiuchi et al.\ take the form
\begin{equation}\label{eq28}
\left(\begin{array}{c} {\bf a}_H\\ {\bf b}_H\\ {\bf c}_H\end{array}\right)=
\left(\begin{array}{ccc} -1 & 0 & 1\\ 1 & 1 & 1\\ 0 & \frac{11}{2} & \frac{11}{2}\\ \end{array}\right)
\left(\begin{array}{c} {\bf a}_R\\{\bf b}_R\\ {\bf c}_R\end{array}\right)=
\left(\begin{array}{ccc} 1 & 0 & 0\\ 0 & 1 & 0\\ \frac{11}{2} & \frac{11}{2} & -\frac{1}{2}\\ \end{array}\right)
\left(\begin{array}{c} {\bf a}_L\\{\bf b}_L\\ {\bf c}_L\end{array}\right)
\end{equation}
and coordinates given in the $L$-system can directly be transformed into the $H$-system
\begin{equation}\label{eq27}
\left(\begin{array}{c} x\\ y\\ z\end{array}\right)_{\!\!\displaystyle H}=
\left(\begin{array}{ccc} 1 & 0 & 11\\ 0 & 1 & 11\\ 0 & 0 & -2\\ \end{array}\right)
\left(\begin{array}{c} x\\ y\\ z\end{array}\right)_{\!\!\displaystyle L}+
\left(\begin{array}{c} 1/2 \\ 0\\ 1/2\end{array}\right)_{\!\!\displaystyle L}\,.
\end{equation}
The translation vector has to be added because of different choices of the origin in both
coordinate systems. With the help of equation (\ref{eq27}) the positional parameters reported by Canfield
for both the high and the low temperature configuration of V$_6$O$_{11}$ yield the structural input for
the band structure calculation, which is summarized in table \ref{tab2}. In addition to the six
inequivalent vanadium atoms there are eleven oxygen sites. Due to the
inversion symmetry of the crystal lattice the unit cell contains 34 atoms. Further structural
data for metallic V$_6$O$_{11}$ were given by Horiuchi et al.\ \cite{horiuchi76}.
\begin{table}[t!]\begin{center}
\begin{tabular}{|c||c|c|c||c|c|c|}\hline
& \multicolumn{3}{|c||}{High temperature structure} & \multicolumn{3}{c|}{Low temperature structure}\\\hline
Atom &\hspace{0.7cm}$x$\hspace*{0.7cm}&\hspace{0.7cm}$y$\hspace*{0.7cm}&\hspace{0.7cm}$z$\hspace*{0.7cm}&
\hspace{0.7cm}$x$\hspace*{0.7cm}&\hspace{0.7cm}$y$\hspace*{0.7cm}&\hspace{0.7cm}$z$\hspace*{0.7cm}\\\hline
V1 & 0.9931 & 0.4467 & 0.0624 & 0.9986 & 0.4633 & 0.0584\\
V2 & 0.9731 & 0.9264 & 0.0672 & 0.0081 & 0.9568 & 0.0612\\
V3 & 0.9652 & 0.4583 & 0.2422 & 0.9578 & 0.4379 & 0.2506\\
V4 & 0.9560 & 0.9463 & 0.2432 & 0.9362 & 0.9362 & 0.2484\\
V5 & 0.9765 & 0.4689 & 0.4170 & 0.9885 & 0.5086 & 0.4074\\
V6 & 0.9823 & 0.9777 & 0.4152 & 0.0582 & 0.0473 & 0.4036\\\hline
O1 & 0.6888 & 0.4114 & 0.0332 & 0.6584 & 0.3744 & 0.0402\\
O2 & 0.3202 & 0.5570 & 0.0666 & 0.3173 & 0.5650 & 0.0658\\
O3 & 0.6845 & 0.4670 & 0.1090 & 0.6809 & 0.4685 & 0.1098\\
O4 & 0.3736 & 0.6180 & 0.1460 & 0.3464 & 0.5937 & 0.1508\\
O5 & 0.6930 & 0.4060 & 0.2134 & 0.6954 & 0.4057 & 0.2150\\
O6 & 0.3138 & 0.5883 & 0.2468 & 0.3360 & 0.6176 & 0.2396\\
O7 & 0.6697 & 0.4485 & 0.2968 & 0.6683 & 0.4469 & 0.2974\\
O8 & 0.3171 & 0.5339 & 0.3474 & 0.3161 & 0.5435 & 0.3468\\
O9 & 0.7039 & 0.4223 & 0.3898 & 0.6826 & 0.3998 & 0.3940\\
O10 & 0.3180 & 0.5976 & 0.4254 & 0.3510 & 0.6150 & 0.4224\\
O11 & 0.6864 & 0.4689 & 0.4718 & 0.6983 & 0.4710 & 0.4704\\\hline
\end{tabular}
\caption{Atomic positions for the high and low temperature structure of V$_6$O$_{11}$ as
used in the band structure calculations. These data have been determined by Canfield
\cite{canfield90}; the coordinates refer to the primitive translations proposed by Horiuchi et al.\ \cite{horiuchi76}.}
\label{tab2}\end{center}\end{table}
However, we use the data of Canfield.

To allow for an adequate interpretation of the electronic structure results, it is important to
preserve close relation between the representation of the V$_6$O$_{11}$ structure and the parent
rutile structure. We arrange the unit cell of the $L$-system
in a Cartesian coordinate system such that the alignment of the oxygen octahedra
resembles the rutile arrangement; we assume
${\bf a}_L=(-a_1,0,a_3)$, ${\bf b}_L=(a_1,b_2,a_3)$, and ${\bf c}_L=(c_1,c_2,c_3)$.
An application of the elementary relations
$a_L^2=a_1^2+a_3^2$, $b_L^2=a_1^2+b_2^2+a_3^2$, and ${\bf a}_L\cdot{\bf b}_L=a_Lb_L\cos\gamma_L$
immediately yields $a_{1/3}^2=1/2\cdot(a_L^2\mp a_Lb_L\cos\gamma_L)$ and $b_2^2=a_L^2-a_L^2$.
Furthemore, from the expressions ${\bf a}_L\cdot{\bf c}_L=a_Lc_L\cos\beta_L$ as well as
${\bf b}_L\cdot{\bf c}_L=b_Lc_L\cos\alpha_L$ one finds
$c_{1/3}=(b_Lc_L\cos\alpha_L\mp a_Lc_L\cos\beta_L-b_2c_2)/2a_{1/3}$,
which implies a quadratic equation for $c_2$ because $c_L^2=c_1^2+c_2^2+c_3^2$. Solving this
equation completes the calculation of the unit cell in the $L$-system and a transformation
via equation (\ref{eq28}) yields the unit cell in the $H$-system. For
the high temperature structure of V$_6$O$_{11}$ at $298$\,K we find ($A=8.6665$\,$a_B$)
\begin{equation}
{\bf a}_H=A\left(\begin{array}{c} -1.0000\\ \phantom{-}0.0000\\ \phantom{-}0.6417\end{array}\right)\,,\;\;\;
{\bf b}_H=A\left(\begin{array}{c}  1.0000\\ 0.9616\\ 0.6417\end{array}\right)\,,\;\;\;
{\bf c}_H=A\left(\begin{array}{c} -0.0053\\ \phantom{-}5.4629\\ \phantom{-}3.6352\end{array}\right)\,.
\end{equation} 
The low temperature structure at $20$\,K results in ($A=8.6981$\,$a_B$)
\begin{equation}
{\bf a}_H=A\left(\begin{array}{c} -1.0000\\ \phantom{-}0.0000\\ \phantom{-}0.6521\end{array}\right)\,,\;\;\;
{\bf b}_H=A\left(\begin{array}{c}  1.0000\\ 0.9224\\ 0.6521\end{array}\right)\,,\;\;\;
{\bf c}_H=A\left(\begin{array}{c} -0.0074\\ \phantom{-}5.3417\\ \phantom{-}3.7636\end{array}\right)\,.
\end{equation}
The Bohr radius is denoted by $a_B$ and the primitive translations refer to Cartesian coordinates.
The corresponding positional parameters of the vanadium and oxygen atoms are denoted in table \ref{tab2}.

The presented LDA calculations are based on the (scalar relativistic) augmented spherical wave (ASW)
method \cite{williams79,eyert00a}. We adopt the parametrization of the exchange
\begin{table}[t!]\begin{center}
\begin{tabular}{|c||c|c||c|c|}\hline
 & \multicolumn{2}{|c||}{High temperature structure} & \multicolumn{2}{c|}{Low temperature structure}\\\hline
Atom &\hspace{0.6cm}Radius\hspace*{0.6cm}&\hspace{0.6cm}Charge\hspace*{0.6cm}&
\hspace{0.6cm}Radius\hspace*{0.6cm}&\hspace{0.6cm}Charge\hspace*{0.6cm}\\\hline
V1 & 2.1866 & 2.5090 & 2.2303 & 2.6179\\
V2 & 2.1587 & 2.7155 & 2.2278 & 2.8276\\
V3 & 2.2833 & 2.5240 & 2.2158 & 2.5534\\
V4 & 2.2755 & 2.7564 & 2.2248 & 2.7939\\
V5 & 2.3440 & 2.7775 & 2.2248 & 2.5213\\
V6 & 2.3437 & 2.7354 & 2.1856 & 2.6201\\\hline
O1 & 1.9684 & 4.1741 & 1.9423 & 4.2072\\
O2 & 1.7883 & 4.0164 & 1.8240 & 3.9466\\
O3 & 1.9680 & 4.3682 & 1.9699 & 4.4187\\
O4 & 2.0093 & 4.3624 & 1.9420 & 4.2381\\
O5 & 1.7654 & 3.8853 & 1.8219 & 3.8832\\
O6 & 1.8752 & 4.0164 & 1.8201 & 3.9683\\
O7 & 1.8673 & 4.0673 & 1.7943 & 3.9711\\
O8 & 1.8908 & 4.0278 & 1.8394 & 4.0294\\
O9 & 1.8585 & 4.0833 & 1.8568 & 4.0498\\
O10 & 1.8828 & 4.1106 & 1.8121 & 3.9479\\
O11 & 1.9170 & 4.0959 & 1.8166 & 3.8850\\\hline
\end{tabular}
\caption{Radii of the vanadium/oxygen spheres (in $a_B$) as well as calculated LDA valence
charges (V $3d$ or O $2p$) for both the high and the low temperature phase of V$_6$O$_{11}$. }
\label{tab7}\end{center}\end{table}
correlation potential introduced by Vosko, Wilk, and Nusair \cite{vosko80}. The ASW scheme
uses the atomic sphere approximation and models the full crystal potential by means of
spherical symmetric potential wells. Here it is required that the atomic spheres fill the
space of the unit cell. For open crystal structures problems can arise because space filling only due to
atom centered spheres may lead to large overlap. As a consequence, so-called empty spheres, i.e.\ pseudo atoms
without nuclei, have to be introduced. They are used to correctly model the
shape of the crystal potential in large voids. The collection of both physical and empty spheres leads to
an artificial close-packed structure. However, since the potential of the whole set of spheres ought to
represent the crystal potential as exactly as possible it is a challenge to find optimal empty sphere
positions and optimal radii for the real and empty spheres. Here the sphere
geometry optimization algorithm described in \cite{eyert98a} is most efficient.

By adding 35 empty spheres from 18 crystallographically inequivalent classes to the
triclinic unit cell of high temperature V$_6$O$_{11}$ it is possible to keep the linear
overlap of real spheres below 18\%. Simultaneously, the overlap of any pair of real and empty
spheres is smaller than 23\%. In the case of the low temperature structure 40 empty spheres from 21 inequivalent
classes allow for reducing the overlaps to less than 18\% and 23\%,
respectively. Summing up, the unit cell entering the LDA calculation comprises 69 spheres for the high
and 74 spheres in the case of the low temperature structure. The radii of the vanadium and oxygen
spheres are summarized in table \ref{tab7}, which additionally denotes the valence charges
arising from the LDA band structure calculation. For both V$_6$O$_{11}$ structures the basis sets taken
into account in the secular matrix comprise V $4s$, $4p$, $3d$, ($4f$) and O $2s$, $2p$, ($3d$) orbitals.
States given in parentheses enter as tails of other states; for details
on the ASW method see \cite{eyert00a,eyert91}. To complete the basis sets we add empty sphere states, which are
determined with respect to the spacial extensions of the spheres and the total charge occupying
them. The used configurations reach from $1s$, ($2p$) to $1s$, $2p$, $3d$, $4f$, ($5g$). During the
course of the LDA calculation the Brillouin zone is sampled with an increasing number of {\bf k}-points
in the irreducible wedge. In this manner one ensures convergence of the results with respect to the
fineness of the {\bf k}-space grid. For both the high and low temperature calculation
the number of {\bf k}-points was increased from 108 to
256, 864, and 2048. Self-consistency of the
charge density was assumed for deviations of the atomic charges and the total energy of subsequent iterations
less than $10^{-8}$ electrons and $10^{-8}$ Ryd, respectively.

As ${\bf V_4O_7}$ reveals short metal chains (length $n=4$) it is suitable for investigating the electronic
features of the sesquioxide-like chain end sites and their influence on the MIT. Using our results for V$_6$O$_{11}$ 
we can understand the MIT of V$_4$O$_7$. Afterwards we establish important implications for the frequently discussed MITs of
V$_2$O$_3$. As for V$_6$O$_{11}$, the space group of V$_4$O$_7$ is $P\bar{1}$ ($C_i^1$) and the MIT at $250$\,K is
accompanied by a structural distortion. Refinements tracing back to Hodeau and Marezio \cite{hodeau78} placed the vanadium
\begin{table}[t!]\begin{center}
\begin{tabular}{|c||c|c|c||c|c|c|}\hline
 & \multicolumn{3}{|c||}{High temperature structure} & \multicolumn{3}{c|}{Low temperature structure}\\\hline
Atom &\hspace{0.7cm}$x$\hspace*{0.7cm}&\hspace{0.7cm}$y$\hspace*{0.7cm}&\hspace{0.7cm}$z$\hspace*{0.7cm}&
\hspace{0.7cm}$x$\hspace*{0.7cm}&\hspace{0.7cm}$y$\hspace*{0.7cm}&\hspace{0.7cm}$z$\hspace*{0.7cm}\\\hline
V1 & 0.9806 & 0.9383 & 0.1008 & 0.9773 & 0.9299 & 0.1043\\
V2 & 0.9946 & 0.4509 & 0.0964 & 0.9781 & 0.4382 & 0.1008\\
V3 & 0.9662 & 0.9613 & 0.3742 & 0.9508 & 0.9512 & 0.3757\\
V4 & 0.9914 & 0.4905 & 0.3658 & 0.0146 & 0.5021 & 0.3608\\\hline
O1 & 0.6927 & 0.4137 & 0.0514 & 0.6932 & 0.4151 & 0.0512\\
O2 & 0.3237 & 0.5551 & 0.1036 & 0.3191 & 0.5383 & 0.1112\\
O3 & 0.6825 & 0.4643 & 0.1716 & 0.6808 & 0.4658 & 0.1718\\
O4 & 0.3763 & 0.6232 & 0.2254 & 0.3804 & 0.6214 & 0.2254\\
O5 & 0.6893 & 0.4112 & 0.3328 & 0.7018 & 0.4240 & 0.3278\\
O6 & 0.3044 & 0.5683 & 0.3910 & 0.2793 & 0.5257 & 0.4068\\
O7 & 0.6605 & 0.4326 & 0.4700 & 0.6616 & 0.4218 & 0.4728\\\hline
\end{tabular}
\caption{Atomic positions for the high and low temperature structure of V$_4$O$_7$
as used in the band structure calculations. These data trace back to
Hodeau and Marezio \cite{hodeau78}; the coordinates refer to the primitive
translations proposed by Horiuchi et al.\ \cite{horiuchi76}.}
\label{tab3}\end{center}\end{table}
and oxygen sites at the Wyckoff positions (2i): $\pm (x,y,z)$. Above the transition
at 298\,K the authors observed the triclinic parameters $a_A=5.509$\,\AA, $b_A=7.008$\,\AA, $c_A=12.256$\,\AA,
$\alpha_A=95.10^{\circ}$, $\beta_A=95.17^{\circ}$, and $\gamma_A=109.25^{\circ}$. The corresponding values
at low temperatures ($120$\,K) are $a_A=5.503$\,\AA, $b_A=6.997$\,\AA, $c_A=12.256$\,\AA,
$\alpha_A=94.86^{\circ}$, $\beta_A=95.17^{\circ}$, and $\gamma_A=109.39^{\circ}$. This refers
to the unit cell of Andersson and Jahnberg, see equation (\ref{eq25a}) with $n=4$.
The primitive translations of Horiuchi et al.\ are given by
\begin{equation}\label{eq36}
\left(\begin{array}{c} {\bf a}_H\\ {\bf b}_H\\ {\bf c}_H\end{array}\right)=
\left(\begin{array}{ccc} -1 & 0 & 1\\ 1 & 1 & 1\\ 0 & \frac{7}{2} & \frac{7}{2}\\ \end{array}\right)
\left(\begin{array}{c} {\bf a}_R\\{\bf b}_R\\ {\bf c}_R\end{array}\right)=
\left(\begin{array}{ccc} 1 & 0 & 0\\ 0 & 1 & 0\\ 2 & \frac{5}{2} & -\frac{1}{2}\\ \end{array}\right)
\left(\begin{array}{c} {\bf a}_A\\{\bf b}_A\\ {\bf c}_A\end{array}\right)
\end{equation}
and we are able to transform coordinates from the $A$-system into the $H$-system
\begin{equation}\label{eq33}
\left(\begin{array}{c} x\\ y\\ z\end{array}\right)_{\!\!\displaystyle H}=
\left(\begin{array}{ccc} 1 & 0 & 4\\ 0 & 1 & 5\\ 0 & 0 & -2\\ \end{array}\right)
\left(\begin{array}{c} x\\ y\\ z\end{array}\right)_{\!\!\displaystyle A}+
\left(\begin{array}{c} 1/2\\ 1/2\\ 1/2\end{array}\right)_{\!\!\displaystyle A}\,.
\end{equation}
Again, the translation vector has been added to account for different choices of the
origin in the coordinate systems. Using equation (\ref{eq33}) the positional parameters of Hodeau
and Marezio for high and low temperature V$_4$O$_7$ result in the values summarized in table \ref{tab3}, which
enter the LDA calculation. Alternative data for the
crystal structure were given by Horiuchi et al.\ \cite{horiuchi72} and Marezio et al.\ \cite{marezio73}.
Altogether, one obtains four crystallographically inequivalent vanadium and seven
oxygen sites. Due to the inversion symmetry of the lattice the V$_4$O$_7$ unit cell contains 22 atoms.

Using the lattice constants of Hodeau and Marezio we set up the
triclinic unit cell of V$_4$O$_7$ in Cartesian coordinates. Because of the similar definitions
of the $A$ and $L$-system we can proceed analogous with our considerations for V$_6$O$_{11}$, applying the parameters 
$a_A$, $b_A$, $c_A$, $\alpha_A$, $\beta_A$ and $\gamma_A$.
Via equation (\ref{eq36}) we find a unit cell in the $H$-system. For the high
temperature structure at $298$\,K this yields ($A=8.7702$\,$a_B$)
\begin{equation}
{\bf a}_H=A\left(\begin{array}{c} -1.0000\\ \phantom{-}0.0000\\ \phantom{-}0.6396\end{array}\right)\,,\;\;\;
{\bf b}_H=A\left(\begin{array}{c}  1.0000\\ 0.9333\\ 0.6396\end{array}\right)\,,\;\;\;
{\bf c}_H=A\left(\begin{array}{c}  0.0071\\ 3.4280\\ 2.3282\end{array}\right)\,.
\end{equation} 
Moreover, the low temperature configuration at $120$\,K results in ($A=8.7691$\,$a_B$)
\begin{equation}
{\bf a}_H=A\left(\begin{array}{c} -1.0000\\ \phantom{-}0.0000\\ \phantom{-}0.6374\end{array}\right)\,,\;\;\;
{\bf b}_H=A\left(\begin{array}{c}  1.0000\\ 0.9313\\ 0.6374\end{array}\right)\,,\;\;\;
{\bf c}_H=A\left(\begin{array}{c}  0.0054\\ 3.4199\\ 2.3141\end{array}\right)\,.
\end{equation}
These primitive translations are used with the positional parameters denoted in table \ref{tab3}.

Similar to the procedure in the case of V$_6$O$_{11}$ we insert additional augmentation spheres into the
\begin{table}[t!]\begin{center}
\begin{tabular}{|c||c|c||c|c|}\hline
 & \multicolumn{2}{|c||}{High temperature structure} & \multicolumn{2}{c|}{Low temperature structure}\\\hline
Atom &\hspace{0.6cm}Radius\hspace*{0.6cm}&\hspace{0.6cm}Charge\hspace*{0.6cm}&
\hspace{0.6cm}Radius\hspace*{0.6cm}&\hspace{0.6cm}Charge\hspace*{0.6cm}\\\hline
V1 & 2.2002 & 2.7937 & 2.1388 & 2.7457\\
V2 & 2.2349 & 2.6572 & 2.3468 & 2.7529\\
V3 & 2.3224 & 2.7329 & 2.1999 & 2.7430\\
V4 & 2.3433 & 2.7245 & 2.3436 & 2.6778\\\hline
O1 & 1.9931 & 4.3397 & 2.0096 & 4.3259\\
O2 & 1.8278 & 4.0266 & 1.9267 & 4.1770\\
O3 & 1.9883 & 4.4309 & 1.9759 & 4.4115\\
O4 & 2.0465 & 4.3530 & 2.0226 & 4.3440\\
O5 & 1.7994 & 3.9182 & 1.7492 & 3.8328\\
O6 & 1.8993 & 4.0359 & 1.8357 & 3.9669\\
O7 & 1.9164 & 4.1097 & 1.9045 & 4.0616\\\hline
\end{tabular}
\caption{Radii of the vanadium/oxygen spheres (in $a_B$) as well as calculated LDA valence charges
(V $3d$ or O $2p$) for both the high and the low temperature phase of V$_4$O$_7$.}
\label{tab8}\end{center}\end{table}
V$_4$O$_7$ structure to account for its openness and to properly model the crystal potential.
For both the high and the low temperature phase it suffices to apply $13$ crystallographically
inequivalent classes and place altogether 24 empty spheres in the triclinic unit cell.
This keeps the linear overlap of physical spheres below $19$\% and the overlap of any pair of physical and
empty spheres below $24$\%. Unit cells entering the LDA calculation comprise 46 spheres;
table \ref{tab8} denotes the radii of the physical spheres and the calculated
valence charges. All technical details of the calculation are the same as for V$_6$O$_{11}$.

By comparing ${\bf V_7O_{13}}$ to ${\bf V_8O_{15}}$ the influence of the chain center atoms on the electronic
properties of the Magn\'eli phases will be investigated quantitatively in section \ref{sec3}.
Such a comparison is of special interest since the former compound is the only member of the Magn\'eli series,
which does not undergo an MIT. In contrast, V$_8$O$_{15}$ exhibits an MIT
at $\approx70$\,K accompanied by structural distortions. V$_7$O$_{13}$ and both 
V$_8$O$_{15}$ phases crystallize in the triclinic space group $P\bar{1}$ ($C_i^1$) with
the vanadium and oxygen atoms at the Wyckoff positions (2i): $\pm (x,y,z)$.
Canfield \cite{canfield90} reported lattice parameters related to the unit cell of Le~Page and Strobel, see
equation (\ref{eq25}). For V$_7$O$_{13}$ we have $a_L=5.439$\,\AA, $b_L=7.013$\,\AA,
$c_L=37.161$\,\AA, $\alpha_L=67.04^{\circ}$, $\beta_L=57.46^{\circ}$, and $\gamma_L=108.92^{\circ}$.
For the room temperature structure of V$_8$O$_{15}$ these constants amount to $a_L=5.431$\,\AA, $b_L=7.017$\,\AA,
$c_L=42.896$\,\AA, $\alpha_L=66.84^{\circ}$, $\beta_L=57.55^{\circ}$, and $\gamma_L=108.94^{\circ}$, whereas
the low temperature configuration is defined by the values $a_L=10.892$\,\AA, $b_L=6.980$\,\AA,
$c_L=85.907$\,\AA, $\alpha_L=66.83^{\circ}$, $\beta_L=57.38^{\circ}$, and $\gamma_L=108.67^{\circ}$.
The latter unit cell is exceptional because it accounts for a superstructure evolving in V$_8$O$_{15}$
at low temperatures, which complicates the considerations. However, for the high
temperature structures we proceed analogous with the previous analysis of V$_6$O$_{11}$.
A detailed discussion of the structural input for the subsequently presented electronic structure calculations
has been given in \cite{us03c}. We use the positional parameters measured by Canfield. Alternative
crystallographic studies of V$_7$O$_{13}$ at room temperature were performed by Horiuchi et al.\ \cite{horiuchi76}.

\section{Dimerization and localization in ${\bf V_6O_{11}}$}
\label{sec1} 
In this section we analyze the MIT of V$_6$O$_{11}$ \cite{us03}. Changes of the electronic structure at
the transition are discussed in relation to the
structural transformations occuring simultaneously. The study will benefit from our unified
representation of the crystal structures of the Magn\'eli phases as well as of VO$_2$ and V$_2$O$_3$.
We will succeed in grouping the electronic bands of V$_6$O$_{11}$ into states behaving similarly to either the
dioxide or the sesquioxide. Therefore it is possible to analyze the phase transitions of the different
compounds on the basis of a common point of view, which helps us to gain insight into the delicate
interplay of electron-lattice coupling and electronic correlations and their influence on the respective MITs.

\begin{figure}[t!]\begin{center}
\includegraphics[width=10.0cm,clip]{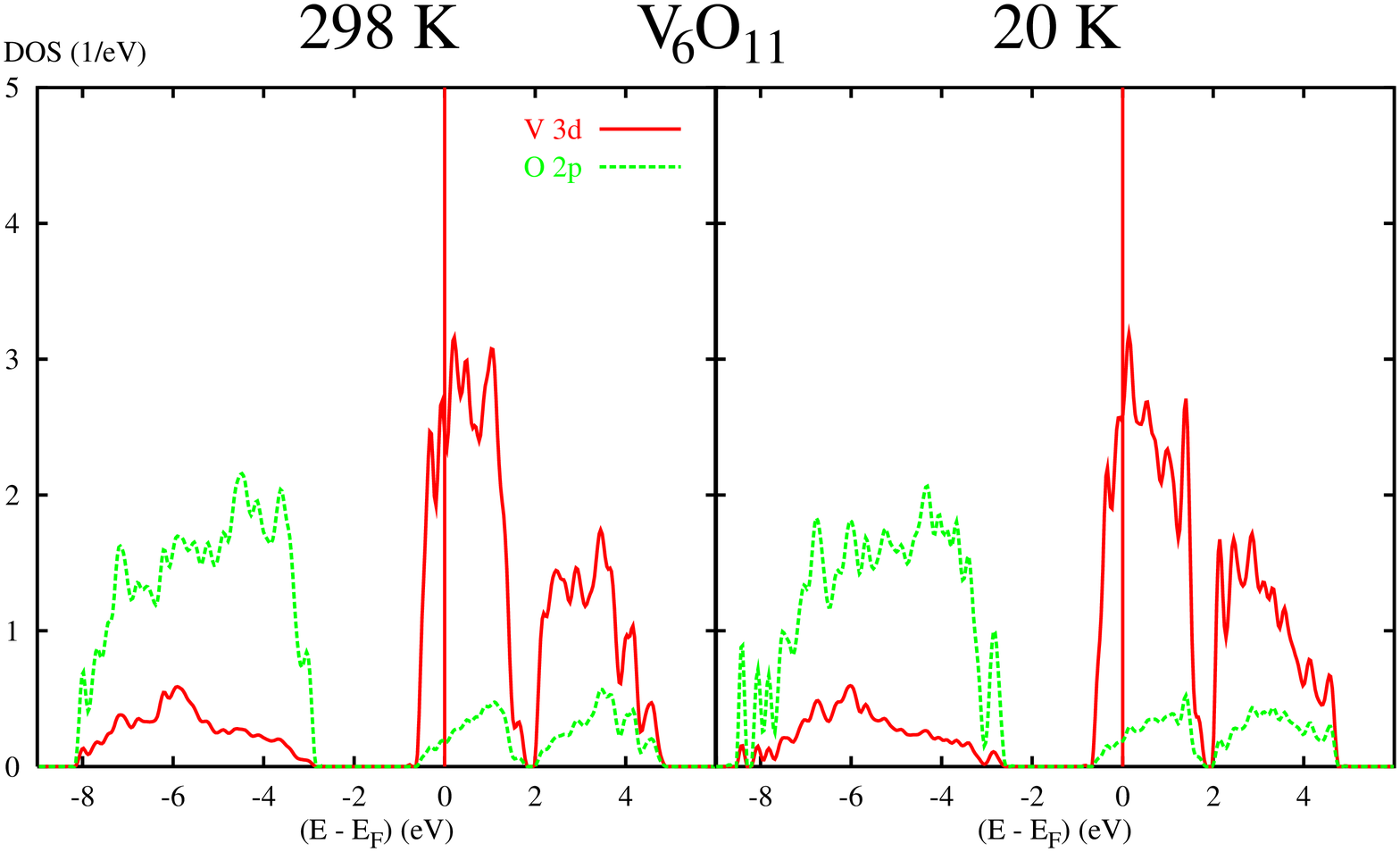}
\caption{Partial V $3d$ and O $2p$ densities of states (DOS) per vanadium atom resulting from the high (298\,K)
and low (20\,K) temperature crystal structure of V$_6$O$_{11}$ are compared.}
\label{pic27}\end{center}\end{figure}

We start our considerations with a survey of the partial densities of states shown in figure \ref{pic27}, which arise from
the LDA calculations for high and low temperature V$_6$O$_{11}$. The low temperature antiferromagnetism is not considered
for the reasons discussed earlier. In the high temperature case we observe three groups of bands in
the energy intervals from $-8.2$\,eV to $-2.9$\,eV, from $-0.7$\,eV to $1.8$\,eV, and from $2.0$\,eV to $4.8$\,eV.
On entering the low temperature phase the lowest group broadens considerably, reaching from $-8.4$\,eV
to $-2.7$\,eV, whereas the other groups are almost not altered. Due to substantial structural similarities to
the dioxide/sesquioxide it is not surprising that the electronic structure of V$_6$O$_{11}$ resembles
the expectations of the molecular orbital picture as discussed for the former compounds.
Characteristic $\sigma$ and $\pi$-type overlap of V $3d$ and O $2p$ states gives rise to bonding-antibonding
split molecular orbitals in the energetical order $\sigma$-$\pi$-$\pi^{*}$-$\sigma^{*}$. Bonding
states are identified with the lowest, $\pi^{*}$ states with the middle, and
$\sigma^{*}$ states with the highest group of bands. Since the metal atoms are octahedrally coordinated, the
$\sigma^{*}$ and $\pi^{*}$ bands display $e_g^{\sigma}$ and $t_{2g}$ symmetry, respectively. Because of 12 V
and 22 O atoms per unit cell there are $12\times2=24$ vanadium $3d$ $e_g^{\sigma}$, $12\times3=36$ vanadium $3d$
$t_{2g}$, and $22\times3=66$ oxygen $2p$ bands. In agreement with the molecular orbital picture
and the findings for VO$_2$ and V$_2$O$_3$ the energetically lowest structure in figure \ref{pic27}
predominantly traces back to O $2p$ states, whereas the other groups mainly originate from V $3d$ states.
Due to a different electron count, the filling of the $t_{2g}$ levels is between that of VO$_2$ and V$_2$O$_3$.
Contributions of vanadium and oxygen in regions dominated by the respective other states are due to $p$-$d$
hybridization. In regions corresponding to $\sigma$-type V-O overlap,
i.e.\ in the $e_g^{\sigma}$ range and in the lower half of the O $2p$ range, they are stronger. Larger $t_{2g}$ admixtures
in the $e_g^{\sigma}$ interval of V$_6$O$_{11}$, compared to VO$_2$, point to increased deviations from an ideal octahedral
coordination. The distortions agree more with those of V$_2$O$_3$.

Differences between the partial densities of states calculated for the high and low temperature structure are small.
No energy gap is observed for the low temperature structure. Instead, 
the structural modifications accompanying the MIT leave the DOS at the Fermi energy almost unaffected. This is
not surprising as for neither the insulating phase of VO$_2$ nor the AFI phase of V$_2$O$_3$ did the
calculations succeed in reproducing an energy gap, which fact was attributed to shortcomings of the LDA. However, as
has been demonstrated for
the dioxide, this does not prevent us from understandig the mechanism of the MIT \cite{eyert02a,wentzcovitch94}.
Since we study the relations between electronic states and local atomic environments as well as the modification of this
relationship at the transition, the LDA limitations do not affect the following considerations.

The structural modifications of V$_6$O$_{11}$ at $170$\,K are mainly concerned with a strong V4-V6
dimerization in the 2-4-6 vanadium chain, see figure \ref{pic19}. While the V4-V6 bond length
decreases from $2.85$\,\AA\ to $2.62$\,\AA, the V2-V4 and V6-V6 distances grow from $2.95$\,\AA\
to $3.06$\,\AA\ and from $2.82$\,\AA\ to $3.30$\,\AA, respectively. As a consequence, two isolated V4-V6
\begin{figure}[t!]\begin{center}
\includegraphics[width=10.0cm,clip]{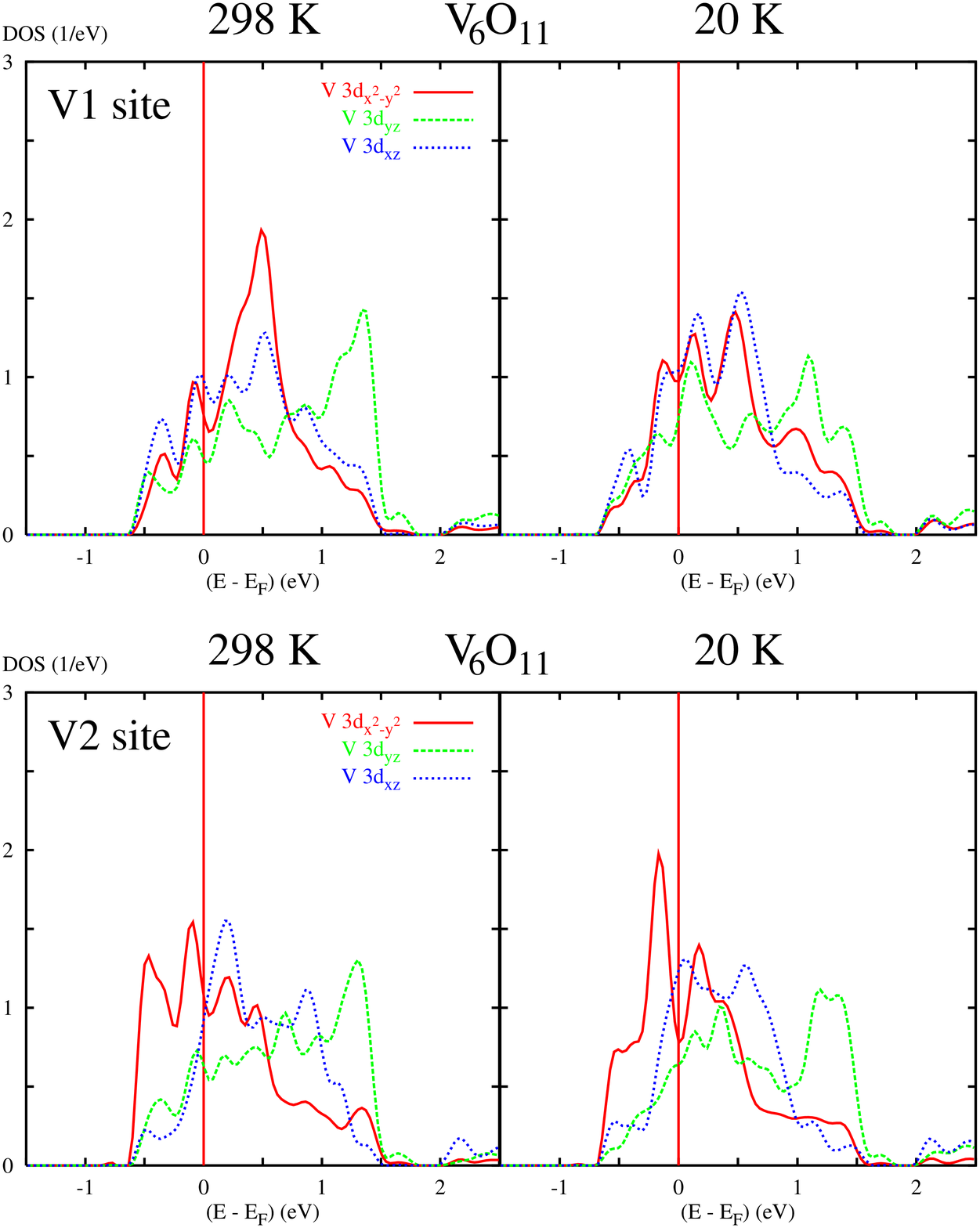}
\caption{Site-projected partial V $3d$ $t_{2g}$ densities of states (DOS) per metal atom for
the high and the low temperature crystal structure of V$_6$O$_{11}$: sites V1 and V2. The crystal structure
is depicted in figure \ref{pic19} and the orbitals refer to the local rotated reference frame.}
\label{pic23}\end{center}\end{figure}
vanadium pairs arise
\begin{figure}[t!]\begin{center}
\includegraphics[width=10.0cm,clip]{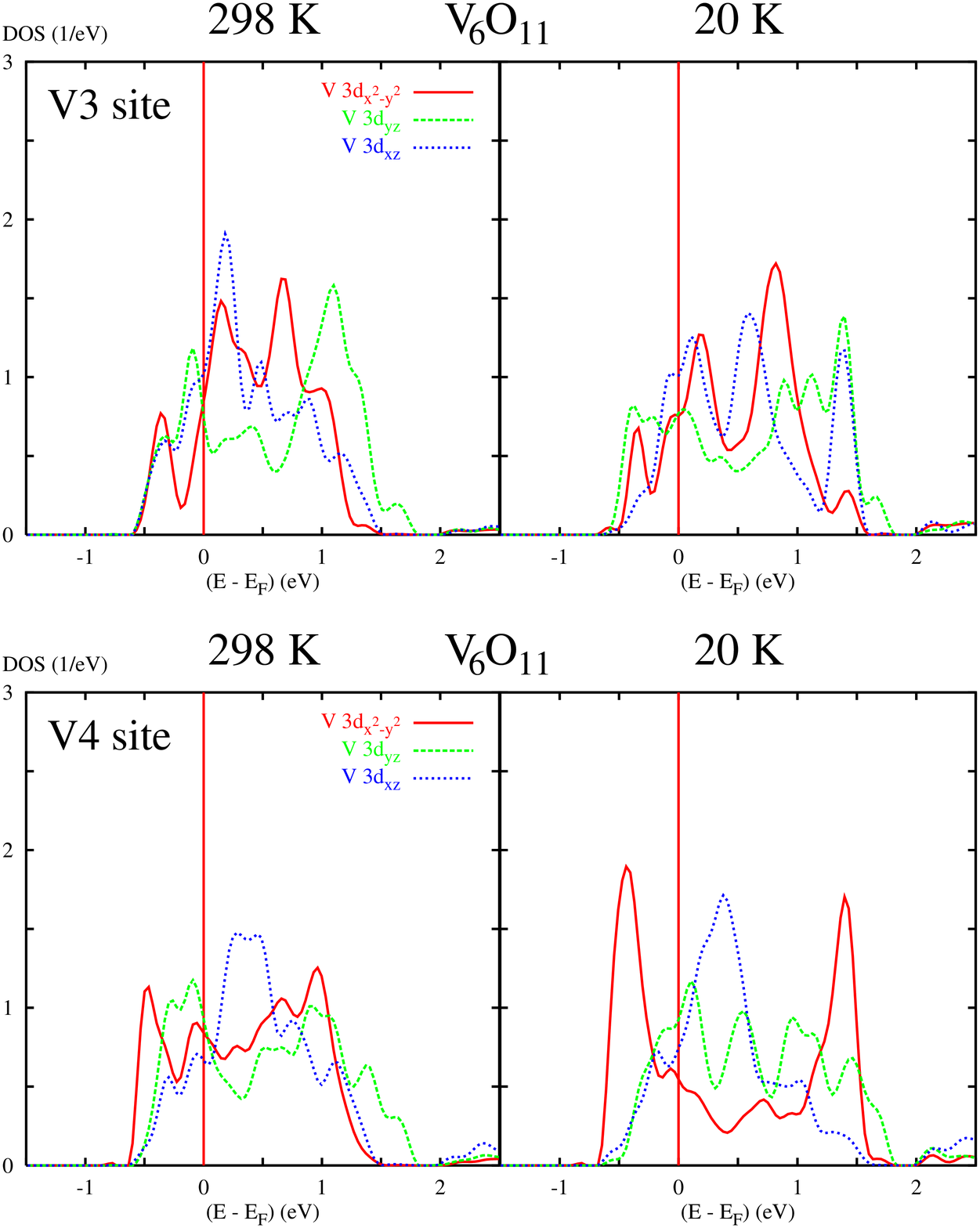}
\caption{Same representation as used in figure \ref{pic23}, but for the vanadium sites V3/V4.}
\label{pic24}\end{center}\end{figure}
and the chain end atoms (V2) separate from the rest of the vanadium chain.
\begin{figure}[t!]\begin{center}
\includegraphics[width=10.0cm,clip]{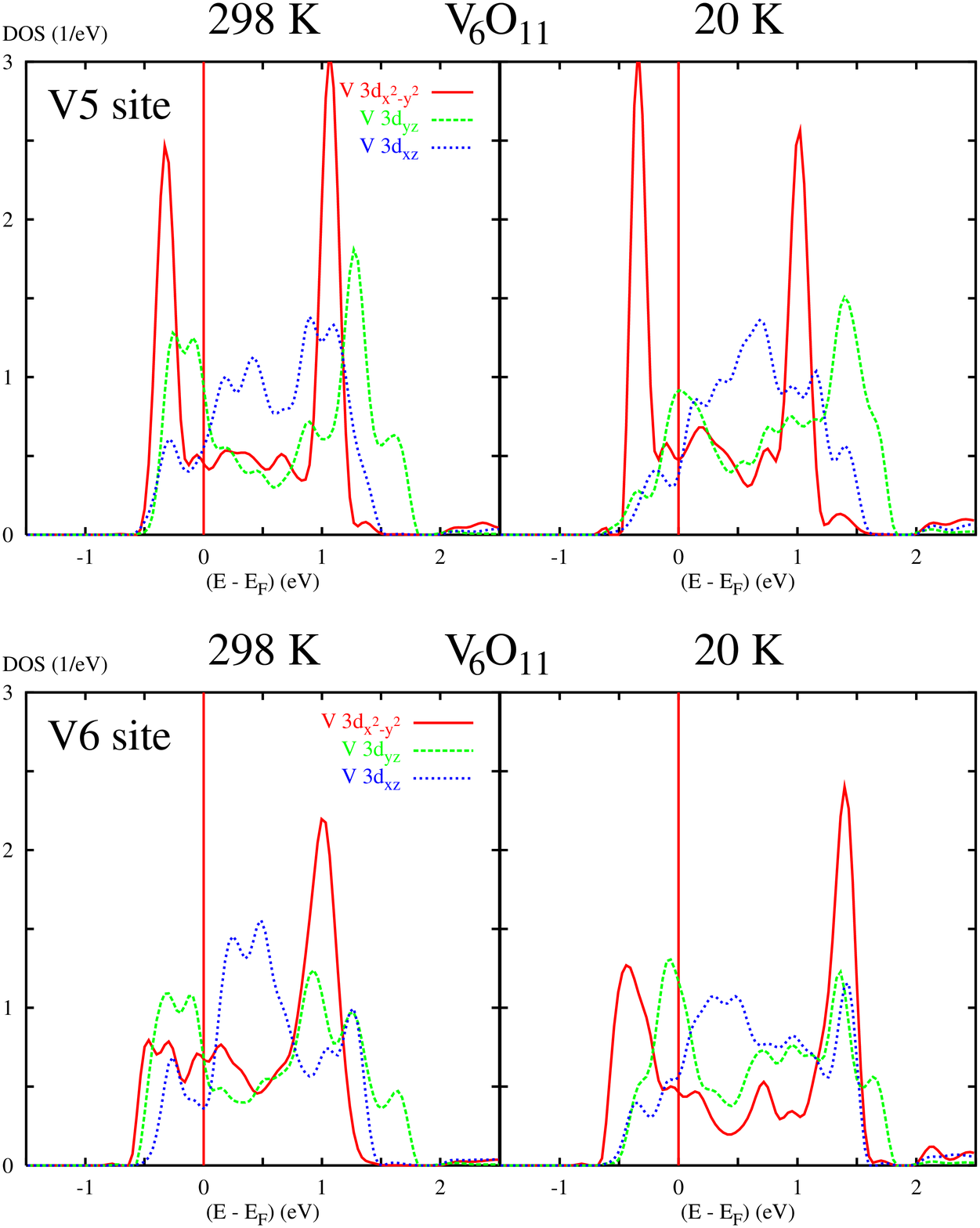}
\caption{Same representation as used in figure \ref{pic23}, but for the vanadium sites V5/V6.}
\label{pic25}\end{center}\end{figure}
In the 1-3-5 chain dimerization effects yield a shortened V5-V5 bond length both above and below the
MIT. Obviously, in the insulating phase the V5-V5 pair separates even more from the evolving V1-V3 pairs. All
these modifications of the crystal structure resemble the pairing effects known from the Peierls
distortion in monoclinic VO$_2$. Due to finite chains in the case of V$_6$O$_{11}$ it is not surprising to find more
complicated distortion patterns; the end atoms affect the V-V pairing in the whole chain.

The monoclinic phase of VO$_2$ is characterized not only by metal-metal pairing but also by
zigzag-type in-plane displacements of the vanadium atoms along the local $z$-axis, thus along
the diagonals of the rutile basal planes. While metal sites in the rutile structure coincide with
the centers of the surrounding oxygen octahedra, they are shifted away from these positions by
$0.20$\,\AA\ in the distorted configuration, thus giving rise to stronger V-O overlap. Equivalent shifts
away from the centers of the octahedra are observed for the sites V3, V4, V5, and V6 in V$_6$O$_{11}$.
The distance between vanadium atoms and octahedral centers grows from $0.16$\,\AA\ to $0.29$\,\AA\
(V3), from $0.17$\,\AA\ to $0.23$\,\AA\ (V4), from $0.07$\,\AA\ to $0.23$\,\AA\ (V5), and from $0.04$\,\AA\ to 
$0.35$\,\AA\ (V6), respectively. Actually, the major part of the shifts is oriented perpendicular to the
$c_{\rm prut}$-axis. Although the metal sites in the high temperature phase do not coincide with
octhedral centers, the relative changes at the MIT are similar to VO$_2$. Interestingly,
the high temperature lateral displacements are larger for the intermediate sites V3 and V4 than for the center
atoms V5 and V6. This is due to a slight rotation of the vanadium chains away from the $c_{\rm prut}$-axis,
which is present in each Magn\'eli phase. Because of the rotation the chain end sites V1 and V2
shift away from their nearest neighbours along $a_{\rm prut}$ -- resembling the
vanadium anti-dimerization of V$_2$O$_3$ parallel to the $c_{\rm hex}$-axis. Here the displacements
of the metal sites with respect to octahedral centers amount to $0.18$\,\AA\ (PM phase) and
$0.21$\,\AA\ (AFI phase). For V$_6$O$_{11}$ these values are $0.32$\,\AA\ in the metallic and $0.35$\,\AA\ in the insulating
phase. Hence, even concerning the details of the local distortion, chain centers and ends behave as their parent
structures VO$_2$ and V$_2$O$_3$.

Next we decompose the V $3d$ $t_{2g}$ group of states into its symmetry
components. Figures \ref{pic23} to \ref{pic25} give site-projected partial
$d_{x^2-y^2}$, $d_{yz}$, and $d_{xz}$ densities of states for the six inequivalent
metal sites V1,...,V6. For each site the presentation of the data refers to the local rotated reference frame,
which was defined in the discussion of the dioxide. The rutile reference frame is useful
since the local octahedral coordinations of the metal sites in V$_6$O$_{11}$ resemble the rutile arrangement,
see the preceding sections.

We turn to the $t_{2g}$ states of the vanadium atoms V4, V5, and V6, which are
involved in strong dioxide-like displacements. Thus the site-projected densities of states are similar
to one another and resemble the DOS of rutile and monoclinic VO$_2$. Note the two peak structure weakly indicated in the
high temperature V4 $d_{x^2-y^2}$ DOS due to $\sigma$-type V-V bonding along the metal chains.
In the low temperature phase the V4-V6 dimerization causes an increased splitting of the DOS
into bonding and antibonding branches located at energies of $-0.5$\,eV and $1.4$\,eV.
The $d_{yz}$ and $d_{xz}$ densities of states undergo energetical upshifts due to
the antiferroelectric displacements of the V4 atoms perpendicular to $c_{\rm prut}$, which raise
the overlap of V $3d$ and O $2p$ states and thus the $\pi$-$\pi^{*}$
splitting. Similar to VO$_2$, the energetical separation of the $d_{x^2-y^2}=d_{\parallel}$ band and the remaining
$e_g^{\pi}$ states is increased but not complete in the low temperature results.

The situation is similar for the V6 atoms, which participate
in the V4-V6 pairs. Therefore the bonding and antibonding branches of the $d_{x^2-y^2}$ DOS appear at
the same energies as for the V4 site. However, in the low temperature configuration the splitting is more
pronounced due to additional V6-V6 bonding. As in the case of V4 the energetical upshift of the
$d_{yz}$/$d_{xz}$ DOS is easily observed, especially directly below the Fermi level. While the
$d_{xz}$ DOS of V4/V6 gives rise to a broad single peak, we find a distinct splitting
of the $d_{yz}$ DOS in contributions centered at about $-0.2$\,eV and $1.0$\,eV for the high
temperature structure. The splitting is understood by taking into account metal-metal
bonding parallel to $b_{\rm prut}$. Across octahedral faces typical sequences V2-V6-V4 appear in this
direction, as obvious from figure \ref{pic9}. Due to a peak in the
low temperature V6 $d_{yz}$ DOS at about $1.4$\,eV, which is not present in the respective V4 DOS but reappears
in the V2 DOS, we can assume increased V2-V6 bonding along $b_{\rm prut}$ at low
temperatures. The $d_{x^2-y^2}$ and $d_{yz}$ densities of states of atom V5 show strong
bonding-antibonding splitting, which hardly changes at the transition. The shape of the
$d_{x^2-y^2}$ DOS is understood in terms of constant V5-V5 bonding. In contrast,
splitting of the $d_{yz}$ DOS traces back to V-V overlap along $b_{\rm prut}$. The
energetical upshift of these states at the MIT is pronounced and has the same origin as discussed for
V4 and V6. In total, the partial V4, V5, and V6 densities of states display characteristic features known from
the dimerization and the zigzag-type displacements in monoclinic VO$_2$ and
are understood in terms of the latter compound.

In contrast to the chain center atoms V4, V5, and V6 the end atoms V1 and V2 are characterized by a
sesquioxide-like local environment. Thus they are involved in V-V bonding across shared
octahedral faces connecting the 1-3-5 to the 2-4-6 layers, but they are not subject to
dimerization along $c_{\rm prut}$. Hence, except for small peaks and shoulders around $-0.5$\,eV
and from $1.0$\,eV to $1.3$\,eV (reminiscent of the chain center $d_{x^2-y^2}$ densities of states)
the V1/V2 $d_{x^2-y^2}$ DOS consists of a single broad
peak extending from $-0.2$\,eV to $0.7$\,eV. The subpeaks observed in this energy region miss counterparts
in the DOS of any neighbouring vanadium atom. Since such peaks are indicative of metal-metal
bonding we conclude, in consistence with the reported V-V distances, that chain end atoms are well
separated from the remainder of the chains. Thus the $d_{x^2-y^2}$ states can be regarded as localized.
While chain center $d_{x^2-y^2}$ orbitals predominantly mediate V-V overlap along $c_{\rm prut}$, this
interaction is suppressed for V1 and V2 for geometrical reasons. Hence the chain end $d_{x^2-y^2}$ states
behave as the $d_{xz}$ states, which generally give rise to reduced ($\pi$-type) metal-metal overlap
along the chains. In both phases the center of weight of the V1 $d_{x^2-y^2}$ DOS is noticeably higher
than that of the V2 $d_{x^2-y^2}$ DOS. Because of nearest neighbours along $a_{\rm prut}$ the
$d_{yz}$ orbitals are important for interpreting the chain end atoms since these states
cause V-V bonding parallel to both $a_{\rm prut}$ and $b_{\rm prut}$. The V1 and V2 $d_{yz}$
states display bonding-antibonding splitting both above and below the transition. For the insulating modification this
separation is more pronounced with peaks directly above the Fermi energy and around $1.2$\,eV. However, so far
we are not able to distinguish between the effects of the different metal-metal interactions.
Whether the shape of the $d_{yz}$ DOS indicates bonding along $a_{\rm prut}$ or $b_{\rm prut}$ or both will be
dealt with later. The question is of interest as the $a_{\rm prut}$-direction of V$_6$O$_{11}$
corresponds to the $c_{\rm hex}$-axis of the corundum structure. Hence the splitting of the $d_{yz}$ states is
equivalent to the $a_{1g}$ splitting in V$_2$O$_3$ and we can gain new insight into the MIT of the sesquioxide. The partial
V1 and V2 $t_{2g}$ densities of states resemble each other due to similar coordinations. They
and are interpreted in accordance with V$_2$O$_3$.

Atom V3 is exceptional as its $t_{2g}$ DOS can neither be interpreted by means of dioxide-like
dimerization in the vanadium chains nor by sesquioxide-like metal-metal overlap along $a_{\rm prut}$.
The second fact is simply due to the absence of adjacent vanadium sites, which allow for overlap between the
1-3-5 and 2-4-6 layers. Nevertheless, all components of the $t_{2g}$ DOS 
are similar to those of sites V1 and V2. The $d_{x^2-y^2}$ and $d_{xz}$ densities of states consist of
broad single peaks. Reminiscent of bonding and antibonding branches in the corresponding V5 DOS we
find satellite peaks at $-0.4$\,eV and a shoulder at $1.0$\,eV in the high temperature phase.
The compact shape of the $d_{xz}$ DOS again is not surprising due to the small V-V overlap this orbital is
involved in, not only for V3 but also for the other vanadium atoms. In contrast, the $d_{yz}$ partial DOS
reveals a distinct double peak structure in both phases, indicative of bonding-antibonding splitting. At first
glance this finding is difficult to understand as geometrically no vanadium partner along $a_{\rm prut}$
exists, which could participate in V-V bonding. However, this puzzling situation is
resolved by reconsidering the possibility of V-V interaction perpendicular to the chains and simultaneously
perpendicular to the $a_{1g}$-direction, namely along $b_{\rm prut}$. As mentioned before, there are
finite vanadium sequences consisting of three atoms (V1-V5-V3), where the distance between neighbouring
atoms corresponds to the longer rutile lattice constant. To be more precise, these bond lengths
amount to $4.58$\,\AA/$4.54$\,\AA\ (V1-V5) and $4.45$\,\AA/$4.40$\,\AA\ (V5-V3) for the high/low temperature
configuration. The associated values in the V2-V6-V4 sequences are $4.54$\,\AA/$4.57$\,\AA\ (V2-V6)
and $4.49$\,\AA/$4.38$\,\AA\ (V6-V4). Since each vanadium atom in V$_6$O$_{11}$
participates in $\sigma$-type overlap along $b_{\rm prut}$, the
bonding-antibonding splitting of the $d_{yz}$ states is present not only for site V3.

This has important implications for the interpretation of the V1/V2 $d_{yz}$ DOS because
interactions parallel to $a_{\rm prut}$ and $b_{\rm prut}$ add to
the bonding-antibonding splitting of these states. The respective antibonding peaks of V1 and V2
are merged at high temperatures but split up at the MIT. Due to the V-V overlap along $b_{\rm prut}$
site V3 resembles the electronic properties of the chain end sites, although its coordination does not
allow for $a_{1g}$-like bonding. Therefore V-V bonding across the vanadium layers is not the only source for
a splitting of the $d_{yz}$ DOS. Despite differing bond lengths hardly any difference between
V-V overlap via octahedral faces within and perpendicular to the layers
is found. Supporting results of Elfimov et al.\ \cite{elfimov03},
V-V couplings other than the $a_{\rm prut}$-like are important for the shape of the $a_{1g}$ bands in V$_2$O$_3$.

While V-O overlap places the V $3d$ $t_{2g}$ states near the Fermi level, the detailed
electronic features of the orbitals, and thus the MIT of V$_6$O$_{11}$, are fundamentally influenced by the
local metal-metal coordination. The small variations of the oxygen
sublattice at the MIT barely affect the shape of any $t_{2g}$ DOS, whereas modifications of the V-V
bond lengths are important. Depending on the coordination, the local electronic properties of the vanadium atoms are
typical for VO$_2$ or V$_2$O$_3$ and it is reasonable to regard the partial
densities of states as local quantities. At the chain centers (V4, V5, V6)
dimerization and antiferroelectric-like displacements via strong electron-lattice interaction cause
a splitting of the $d_{x^2-y^2}$ DOS and an energetical upshift of the $e_g^{\pi}$ states at the MIT.
For atom V5 the former effect is already present at high temperatures. Summarizing,
chain center atoms behave analogous with VO$_2$. In contrast, at the chain ends
(V1, V2, V3) the $d_{x^2-y^2}$ orbitals are characterized by strongly reduced metal-metal overlap and
show a localized nature. The shape of the $d_{x^2-y^2}$ DOS is closely related to
the $d_{xz}$ DOS. Due to the strong localization, electronic correlations are important
for the chain end $d_{x^2-y^2}$ states. Being subject only to in-layer V-V overlap,
the partial V3 $d_{yz}$ DOS shows bonding-antibonding splitting similar to the atoms V1 and V2,
which have an additional nearest neighbour across the layers. Therefore the in-plane V-V interaction is at least as
important as the perpendicular overlap. In conclusion, the sesquioxide-like regions of the V$_6$O$_{11}$ crystal
appear to be susceptible to electronic correlations for geometrical reasons. Structural variations at the
MIT leave all three components of the $t_{2g}$ partial DOS almost unchanged, thus reflecting
close relations to V$_2$O$_3$. We interpret the phase transition of V$_6$O$_{11}$ as resulting
from a combination of electron-lattice interaction and electronic correlations. Due to the
different atomic arrangements in VO$_2$ and V$_2$O$_3$-like regions neither an embedded
Peierls instability nor correlations can account for the MIT of V$_6$O$_{11}$ on their own.

Going back to the calculated valence charges at the vanadium and oxygen sites summarized in table
\ref{tab7} we investigate possible charge ordering. Different atomic spheres radii, necessary to fulfill the requirements
of the atomic sphere approximation, prohibit a quantitative comparison of the
valence charges but allow still for qualitative results. For the
high temparature modification there are about $0.2$ electrons less at sites V1 and V3
than at sites V2 and V4, whereas the charges at the central sites V5 and V6 coincide.
The atomic spheres comprise almost the same space for the pairs V1/V2, V3/V4, and V5/V6, respectively.
In general, the calculated $2.5$/$2.7$ V $3d$ valence electrons contradict the simple ionic picture
predicting an average of $1.33$ electrons per vanadium atom. Instead of
V$^{3.67+}$ ions we find V$^{2.3+}$ and V$^{2.5+}$ configurations, which is not
surprising due to covalent V-O bonding. Oxygen spheres comprise about four $2p$ electrons. Further
electrons enter the empty spheres and hence the open space between vanadium and oxygen.

The accuracy of comparing calculated valences is limited as the assignment
of charge to specific sites is arbitrary to some degree. We assign
charge within an atomic sphere to the respective atomic site. Electrons entering interstitial
empty spheres cannot be assigned to any specific atom.
Importantly, differences in the occupation of equivalent states can be identified with a much higher accuracy.
Thus the $3d$ charge of both V1 and V3 is significantly reduced.
In addition, we obtain the same result for the low temperature phase, where
all vanadium spheres show similar radii. The valence charge at sites V1, V3, and V5 is roughly $0.2$
electrons smaller than at sites V2, V4, and V6, resulting in a charge
transfer of $0.1$ electrons per site from the 1-3-5 to the 2-4-6 chain above and below the
MIT. The different electron count in the two metal chains is related to their
crystallographic inequivalence. While donor atoms (to some extend) approach the VO$_2$ charge
configuration, the acceptor chain resembles more the V$_2$O$_3$ filling. The electronic configurations
of VO$_2$ and V$_2$O$_3$ seem preferable to a mixture of both. However,
chain end sites do not generally gain excess charge at the expense of the dioxide-like chain
centers. Hence it is not the electron count of a specific atom but the local coordination which
determines the behaviour at the phase transition. The calculated valence charges contradict V$^{3+}$/V$^{4+}$ states, but
confirm two kinds of metal valences; they point to smaller charge differences. Deviations
from the ionic picture are consistent with susceptibility data by Gossard et al.\ \cite{gossard74a}, whereas for the
insulating phase we do not find the proposed charge ordering.

The above findings have important implications for understanding V$_2$O$_3$ as the
chain end atoms of the Magn\'eli phases resemble the local metal coordination of this material. According to
the previous discussion the electronic properties of V$_6$O$_{11}$ are influenced by essentially
three effects. First, the center atoms of the vanadium chains along $c_{\rm prut}$ reflect
characteristics of the embedded Peierls instability responsible for the MIT in VO$_2$. Second, the
localization of the chain end $d_{x^2-y^2}$ orbitals leaves them susceptible to electronic
correlations. Third, the chain end $d_{yz}$ orbitals to roughly equal parts are subject to both metal-metal
bonding within (along $b_{\rm prut}$) and perpendicular to (along $a_{\rm prut}$) the vanadium layers. As
vanadium chains in the case of V$_2$O$_3$ ($n=2$) degenerate to pairs, a dioxide-like Peierls instability
cannot contribute as a matter of principle. However, the localized nature of the $d_{x^2-y^2}$ orbitals
and the importance of strong V-V coupling along $b_{\rm prut}$ should be transferred to the sesquioxide.
Due to the high crystal symmetry both effects could not be analyzed in previous studies of V$_2$O$_3$.
As a consequence of the broken symmetry in the Magne\'li phases, combined with a closely related coordination
of appropriate vanadium atoms, they are accessible to the study of V$_6$O$_{11}$. The splitting of the $a_{1g}$-like
states due to metal-metal overlap along $c_{\rm hex}$ therefore seems to be less important than commonly assumed
for V$_2$O$_3$.

\section{Band narrowing in ${\bf V_4O_7}$}
\label{sec2}
The compound V$_4$O$_7$ is suitable for studying chain end effects
since the metal chains here comprise only four atoms \cite{us03b}.
Analogous with the preceding section we discuss the electronic structure and its changes at the MIT
in relation to the simultaneous structural transformations. Applying the knowledge established
in the study of V$_6$O$_{11}$ we can group the electronic bands into states behaving either VO$_2$ or V$_2$O$_3$-like.

\begin{figure}[t!]\begin{center}
\includegraphics[width=10.0cm,clip]{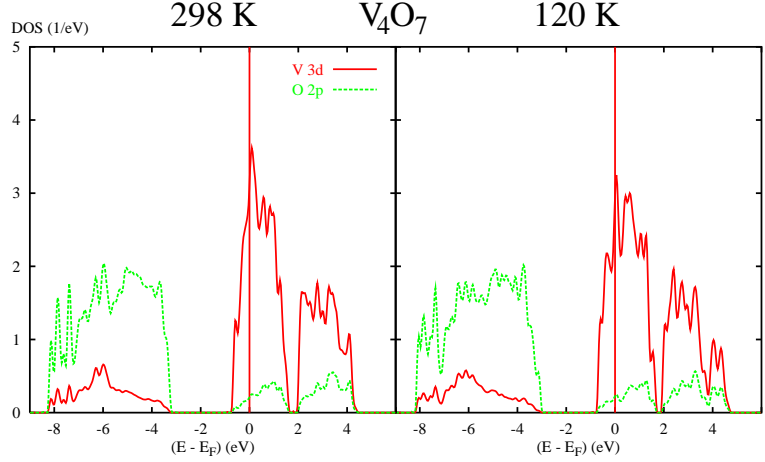}
\caption{Partial V $3d$ and O $2p$ densities of states (DOS) per vanadium atom resulting from
the high (298\,K) and low (120\,K) temperature structure of the compound V$_4$O$_7$.}
\label{pic26}\end{center}\end{figure}

Partial V $3d$ and O $2p$ densities of states arising from the high and low temperature structure are shown in
figure \ref{pic26}. As for VO$_2$, V$_2$O$_3$, and V$_6$O$_{11}$, we find three separated groups of electronic bands.
Recalling the molecular orbital picture, the bonding-antibonding splitting due to V-O overlap, and the discussions in
the previous sections, we identify the bonding states with those bands extending from $-8.3$\,eV to
$-3.3$\,eV at high temperatures; they shift to the region from $-8.2$\,eV to $-3.0$\,eV below the MIT.
Further groups of bands correspond to $\pi^{*}$ and $\sigma^{*}$
states. Hence they are dominated by vanadium states and separate into $t_{2g}$ and $e_g^{\sigma}$
contributions due to the octahedral coordination of the metal sites. While the $t_{2g}$ group
is observed between $-0.7$\,eV and $1.6$\,eV in the metallic phase, its upper edge shifts to
$1.7$\,eV at low temperatures. The $e_g^{\sigma}$ group starts at $2.0$\,eV in both phases,
but it extends to $4.7$\,eV instead of $4.4$\,eV in the insulating phase. Since the unit
cell of V$_4$O$_7$ contains 8 vanadium and 14 oxygen atoms we have $8\times2=16$ vanadium $3d$ $e_g^{\sigma}$,
$8\times3=24$ vanadium $3d$ $t_{2g}$, and $14\times3=42$ oxygen $2p$ bands. Hybridization between V $3d$ and O $2p$
orbitals results in finite contributions of vanadium in the bonding and of oxygen in the antibonding region.
As expected, such hybridization effects are stronger for $\sigma$-type V-O overlap.
Moreover, there is a finite $t_{2g}$ DOS in the $e_g^{\sigma}$ dominated region.
As the magnitude of this admixture equals the result for V$_6$O$_{11}$, deviations from an ideal octahedral 
coordination of the vanadium sites are similar in both compounds and likewise similar to the sesquioxide.
All these results for V$_4$O$_7$ fit very well into the picture developed in the preceding sections.
Due to the formal V $3d$ charge of $1.5$ electrons per metal
site, V$_4$O$_7$ systematically is between V$_6$O$_{11}$ and V$_2$O$_3$. Therefore the V $3d$
$t_{2g}$ orbitals are filled to a somewhat larger extent than for
V$_6$O$_{11}$, which is confirmed by figures \ref{pic27} and \ref{pic26}. Although the calculation
misses an energy gap for the experimental insulating phase,
the LDA shortcomings again do not affect an investigation of relations between
structural and electronic properties or the analysis of implications for the MIT.

\begin{figure}[t!]\begin{center}
\includegraphics[width=10.0cm,clip]{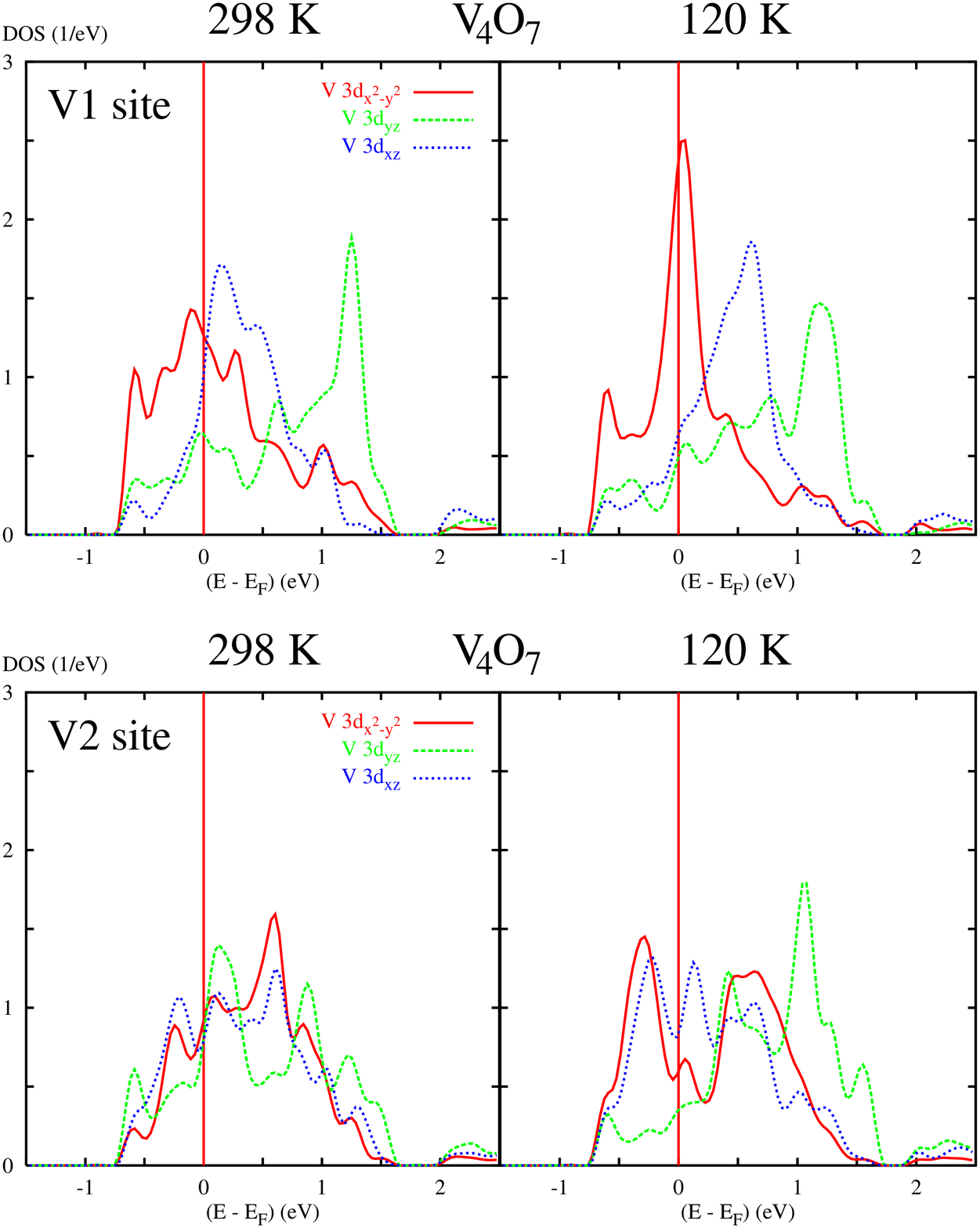}
\caption{Site-projected partial V $3d$ $t_{2g}$ densities of states (DOS) per metal atom for
the high and the low temperature crystal structure of V$_4$O$_7$: sites V1 and V2. The crystal structure
is depicted in figure \ref{pic20} and the orbitals refer to the local rotated reference frame.}
\label{pic17}\end{center}\end{figure}

\begin{figure}[t!]\begin{center}
\includegraphics[width=10.0cm,clip]{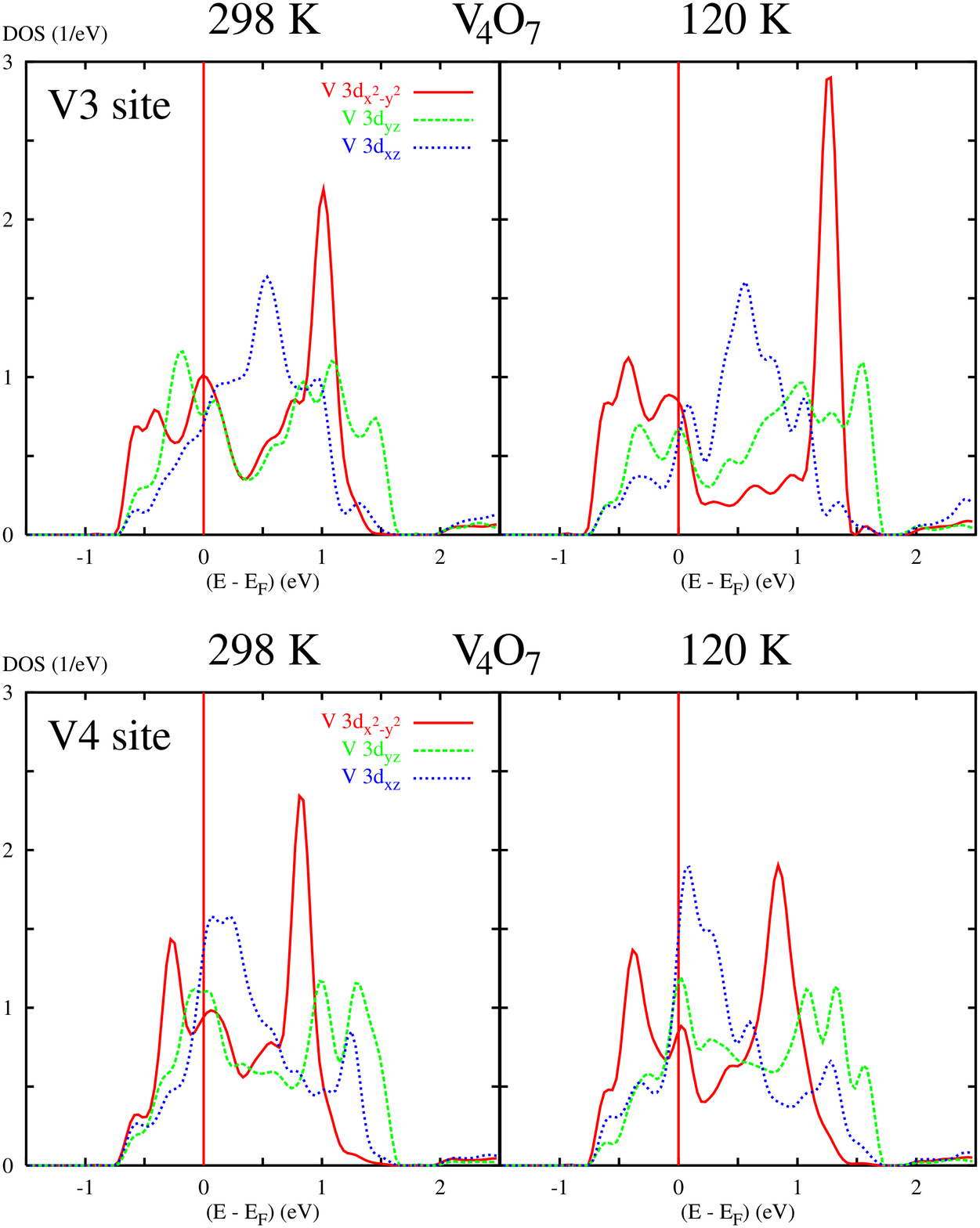}
\caption{Same representation as used in figure \ref{pic17}, but for the vanadium sites V3/V4.}
\label{pic18}\end{center}\end{figure}

While the previous analysis of V$_6$O$_{11}$ gave strong indications for a predominant influence of
electron-lattice interaction in dioxide-like and of electronic correlations in sesquioxide-like
regions of the crystal, we expect that the investigation of the sesquioxide-related member V$_4$O$_7$ will allow for
conclusions about the MIT of V$_2$O$_3$. Separated by oxygen atoms, inequivalent vanadium layers alternate along the
$a_{\rm prut}$-axis of V$_4$O$_7$, comprising the atoms V1/V3 or V2/V4.
Due to this alternation and due to relative shifts of the metal 4-chains along
$c_{\rm prut}$ the end sites V1 and V2 are located on top of each other, see figure \ref{pic20}. We again refer the
V $3d$ orbitals to local coordinate systems with the $z$-axis oriented parallel
to the apical axis of the local octahedron and the $x$-axis parallel to $c_{\rm prut}$, respectively.
Using this reference frame, a decomposition of the V $3d$ $t_{2g}$ group of states into the symmetry components
$d_{x^2-y^2}$, $d_{yz}$, and $d_{xz}$ is depicted in figures \ref{pic17} and \ref{pic18}. The site-projected
partial densities of states are given for the vanadium sites
V1, V2, V3, and V4. In the following, we concentrate on the results for the 1-3 chain, which is
motivated by the fact that the modifications of the local surroundings of atom V1 at the MIT are similar to those
known from V$_2$O$_3$. According to the data denoted in figure \ref{pic20} the V1-V2 and V1-V3 distances
increase at the MIT, as do the distances $a$ and $b$ in the sesquioxide at the PM-AFI transition, see figure
\ref{pic10}. Since in the latter compound all vanadium atoms are crystallographically
equivalent a complete symmetry analysis of the electronic states cannot be performed. However, this is
possible for V$_4$O$_7$. As demonstrated in the case of V$_6$O$_{11}$, the local atomic arrangements
are related to the electronic structures of the respective sites. Therefore the V1 site of
V$_4$O$_7$ allows us to study the influence of the structural changes at the MIT of V$_2$O$_3$
in more detail than in a direct investigation.

Changes in the V$_4$O$_7$ crystal structure at 250\,K are mainly concerned with strong dimerization
evolving in both the 1-3 and 2-4 metal chain, see figure \ref{pic20}. While the V3-V3 and V2-V4 bond
lengths shorten from $2.79$\,\AA\ and $2.93$\,\AA\ to $2.67$\,\AA\ and $2.83$\,\AA, the remaining
distances V1-V3 and V4-V4 elongate from $2.97$\,\AA\ and $2.93$\,\AA\ to $3.03$\,\AA, respectively.
Hence we find an isolated V3-V3 pair in the first and two V2-V4 pairs in the second chain.
In the 1-3 chain the end atoms decouple to a large degree from the central V3-V3 pair. The
modifications resemble pairing effects associated with the embedded Peierls instability of the dioxide. Nevertheless,
the end atoms of the 1-3 chain likewise show structural changes known from the symmetry
breaking PM-AFI transition of V$_2$O$_3$. The distance between V1 and the nearest
vanadium site along $c_{\rm prut}$ increases by $0.06$\,\AA, which is half the
shift found for the sesquioxide. Note that the dominating structural changes in V$_2$O$_3$
do not appear along $a_{\rm prut}$ but along $c_{\rm prut}$. In contrast to the 1-3 chain, the end
atoms of the 2-4 chain increase the coupling to adjacent in-chain sites below the MIT.
Thus they show the opposite trend than found in V$_2$O$_3$ and are not suitable
for investigating the latter material.

Beyond metal-metal pairing also zigzag-type in-plane displacements of the metal sites are present
in the V$_4$O$_7$ structure. The distances between vanadium atoms and octahedral centers amount to
$0.31$\,\AA/$0.35$\,\AA\ (V1), $0.32$\,\AA/$0.27$\,\AA\ (V2),
$0.15$\,\AA/$0.17$\,\AA\ (V3), and $0.08$\,\AA/$0.09$\,\AA\ (V4) in the high/low
temperature modification. Apparently, the average bond length does not increase at the
transition, which contrasts findings for VO$_2$ and V$_6$O$_{11}$, hence confirming the
sesquioxide-like character of V$_4$O$_7$. Because of a slight rotation of the vanadium chains away from the
$c_{\rm prut}$-axis, the lateral displacement is much larger for the chain end sites V1 and V2. The rotation
enables the latter sites to increase their distance along $a_{\rm prut}$, which reflects the
anti-dimerization of V$_2$O$_3$ parallel to $c_{\rm hex}$. In general, the main part of the
zigzag-type displacements is oriented perpendicular to $c_{\rm prut}$. Atom V2 shifts almost
along the apical axis of its octahedron, i.e.\ along the local $z$-axis, whereas atom V1
moves perpendicular to the local $z$-axis. However, because of the rotation of the local coordinate system by
$90^{\circ}$ around the $c_{\rm prut}$-axis, the displacements occur in every chain along the same direction.
On closer inspection we recognize the motions to be antiparallel, which
increases the in-pair V1-V2 bond length. Altogether, the
zigzag-type antiferroelectric mode of the 2-4 chain resembles the behaviour of the dioxide,
whereas the 1-3 chain reveals distinct differences above as well as below the MIT. In retrospect, this
interrelation applies also to V$_6$O$_{11}$ because the displacements of the dioxide-like 1-3-5 chain
point along the local $z$-axis rather than those of the 2-4-6 chain.

Decomposing the V $3d$ $t_{2g}$ partial DOS, we observe in figures \ref{pic17} and \ref{pic18} for the
$d_{x^2-y^2}$ states a similar
behaviour as in VO$_2$. Due to strong $\sigma$-type V-V bonding along the
vanadium chains the chain center sites show distinct splitting of the $d_{x^2-y^2}$ DOS. In the
case of atom V3 a broad peak is observed from roughly $-0.7$\,eV to $0.3$\,eV and a sharp peak
at $1.0$\,eV. Because of weaker V4-V4 bonding the splitting is
smaller for the V4 site with peaks at approximately $-0.3$\,eV and $0.8$\,eV. The chain end atoms V1 and V2
are less affected by in-chain metal-metal bonding and consequently display a more compact $d_{x^2-y^2}$
DOS. At low temperatures the behaviour of both metal chains changes substantially. Increased V-V
dimerization yields enhanced bonding-antibonding splitting of the V3 $d_{x^2-y^2}$ DOS.
Additionally, by virtue of reduced V1-V3 overlap along $c_{\rm prut}$ the V1 $d_{x^2-y^2}$ states
tend to localize. Accordingly, the V1 $d_{x^2-y^2}$ DOS sharpens and develops a pronounced peak near the Fermi
energy at low temperatures. While the V4-V4 distance grows by
$0.1$\,\AA\ below the transition, the coupling of the V4 atoms to the chain end sites increases.
The reduced V2-V4 bond length causes stronger bonding-antibonding splitting of the V2 $d_{x^2-y^2}$ DOS.

Modifications of the vanadium $d_{xz}$ partial DOS at the transition are less significant. They consist
mainly of an energetical up and downshift of the states as observed for atoms V1 and V2, respectively.
Such shifts are due to increased and decreased displacements perpendicular to
$c_{\rm prut}$. Moving vanadium atoms laterally off the centers of gravity of their oxygen
octahedra modifies the overlap between V $3d$ and O $2p$ orbitals. The oxygen sublattice
itself is hardly affected by the structural transformations. Therefore the shifts of the $d_{xz}$ states
can be understood in terms of a zigzag-type antiferroelectric distortion of the vanadium chains.
As the distortion is reduced for the chain center sites V3 and V4, see the former discussion, the
effects are barely visible in the DOS. In the prototypical case of VO$_2$ lateral displacements
are found only in the low temperature monoclinic phase, whereas V$_4$O$_7$ reveals them similar at all temperatures.
Since metal-metal bonding is less important for $d_{xz}$ orbitals no bonding-antibonding splitting is visible.

The situation is more complicated for $d_{yz}$ orbitals, which are considerably influenced
by different kinds of hybridization. Recall from our earlier investigations that these electrons can be
involved in metal-metal overlap across octahedral faces both along $a_{\rm prut}$ and $b_{\rm prut}$.
Moreover, antiferroelectric-like displacements of the vanadium atoms lead to increased overlap between
V $3d$ and O $2p$ states, which causes energetical upshift of the antibonding vanadium dominated
states. In the case of V$_4$O$_7$ the atoms V3 and V4 display neither relevant antiferroelectric
shifts nor V-V interaction parallel to the $a_{\rm prut}$-axis. There are simply no nearest metal
neighbours in this direction. As a consequence, the pronounced double peak structure of the V3/V4
$d_{yz}$ DOS indicates bonding-antibonding splitting due to V1-V3 and
V2-V4 overlap, respectively, which is true at high and low temperatures.
Analogous with the finite metal 3-chains of V$_6$O$_{11}$ along $b_{\rm prut}$ we are confronted with
vanadium pairs giving rise to V1-V3 and V2-V4 bonding across shared octahedral faces. In contrast to the
central vanadium atoms the chain end atoms V1 and V2, apart from participating in V1-V3/V2-V4 overlap
along $b_{\rm prut}$, take part in V1-V2 bonding along $a_{\rm prut}$. The partial V1 and V2 $d_{yz}$
densities of states thus consist of features reminiscent of the sites V2/V3 and V1/V4, respectively. Being
subject to two different types of metal-metal overlap, the $d_{yz}$ orbitals of
the chain end atoms undergo two intertwining bonding-antibonding splittings, which lead to
complicated triangular DOS shapes, especially in the low temperture phase, and
resemble the shape of the V $3d$ $a_{1g}$ DOS of V$_2$O$_3$. In addition, the form is affected by the discussed
zigzag-type displacements of the chain end atoms, which intensify the metal-oxygen bonding.
As a consequence, the center of gravity of the V1/V2 $d_{yz}$ DOS shifts to
higher energies below the MIT. Changes in the shape of the V2 $d_{yz}$ DOS yield a closer
resemblance to the V1 $d_{yz}$ DOS, which can be traced back to an improved coupling
between the chain end atoms at low temperatures. As the relative displacements of the sites V1 and V2 along
$c_{\rm prut}$ are reduced, the V1-V2 bond penetrates the octahedral faces along $a_{\rm prut}$ more
favorably and the overlap becomes more effective.

In conclusion, the electronic structure of V$_4$O$_7$ is strongly influenced by the coordination
of the metal sites, as previously reported for V$_6$O$_{11}$. Generally, while the overlap
of V $3d$ and O $2p$ orbitals places the V $3d$ $t_{2g}$ states close to the Fermi energy, the details of the
electronic structure are subject to metal-metal bonding. In particular, one finds a
distinct bonding-antibonding splitting of the $d_{x^2-y^2}$ states at the chain center sites. In the
low temperature phase the same applies to atom V2. All these splittings are analogous with
the corresponding orbitals in VO$_2$ and V$_6$O$_{11}$, where they likewise are enforced
by V-V dimerization. In contrast, the V1 $d_{x^2-y^2}$ DOS reveals a compact structure and a distinct sharpening
below the MIT as these states hardly take part in the dimerization along the metal chains.
Hence they may be susceptible to electronic correlations. For the $d_{xz}$ and $d_{yz}$ states
we find noticeable response to the displacements of the metal atoms perpendicular to $c_{\rm prut}$.
As the atoms move relative to the octahedral centers, states with $d_{xz}$ and $d_{yz}$ symmetry
display energetical up and downshifts in the vanadium and oxygen dominated regions, respectively.
However, the $d_{yz}$ DOS is affected mainly by the
V-V bonding across shared octahedral faces. Chain center sites are subject to bonding only along
$b_{\rm prut}$ and show a characteristic two peak structure. In contrast, atoms V1 and V2 additionally
take part in $d_{yz}$-type overlap along $a_{\rm prut}$, leading to densities of states similar to
the $a_{1g}$ DOS of V$_2$O$_3$. The latter orbital mediates V-V bonding along $c_{\rm hex}=a_{\rm prut}$.

Similar to the V$_6$O$_{11}$ findings, the LDA results indicate small charge transfer between the two
inequivalent metal chains. Recall the limitations of assigning charge to specific atomic sites as discussed
in section \ref{sec1}. According to table \ref{tab8} the V2 sites comprise $0.2$ electrons less than the
V1 sites in the high temperature modification of V$_4$O$_7$, while V3 and V4 show
similar values. Note that in the pairs V1/V2 and V3/V4 the sphere radii are almost equal. The interpretation
of the low temperature results is a little more complicated. However, taking into account the various sphere radii
one can state reduced charges for both V2 and V4, where the charge deficit is 
$0.2$ electrons per site. Of course, the calculated values of $2.6$-$2.8$ V $3d$ valence electrons do not agree
with the ionic picture predicting $1.5$ electrons per metal atom -- but due to
covalent V-O bonding this deviation is not surprising. In comparison to V$_6$O$_{11}$
the atomic spheres of V$_4$O$_7$ comprise slightly more charge, which reflects the position of the compound
in the Magn\'eli series. We find no significant differences in the charges of the particular oxygen spheres
since each of them contains about four $2p$ electrons. Further electrons enter the interstitial
empty spheres and thus are difficult to classify in vanadium and oxygen-type contributions. Altogether we observe
a charge transfer of about $0.1$ electrons per site from the 2-4 to the 1-3 chain. In the high temperature phase
only the chain end sites are involved in the transfer. As in the case of V$_6$O$_{11}$ the different electron count
is related to the crystallographical inequivalence of the two vanadium chains. While the 2-4
donator chain becomes slightly more VO$_2$-like, the 1-3 acceptor chain to some extent approaches the V$_2$O$_3$
configuration. These LDA findings qualitatively confirm x-ray results by Marezio et al.\ \cite{marezio72},
who applied an ionic picture and reported chains of V$^{3+}$ and V$^{4+}$ ions running parallel to the
$c_{\rm prut}$-axis. Although an ordered arrangement of varying valence charges is confirmed by the calculation,
the charge discrepancy is much smaller due to covalent bonding. Nuclear magnetic resonance findings by
Gossard et al.\ \cite {gossard74} support chains of differently charged atoms along $c_{\rm prut}$ but
indicate incomplete differentiation into 3+ and 4+ valences. The authors observed a slightly increased charge
ordering below the MIT, which is confirmed by the LDA results.

As in other Magn\'eli phases, particularly in V$_6$O$_{11}$, the electronic
structure of V$_4$O$_7$ is a mixture of features characteristic of either vanadium dioxide or
sesquioxide. The local atomic environment of the chain end site V1 bears close resemblance to the
V$_2$O$_3$ structure as it is subject to an increased separation from its
vanadium neighbours at the MIT. We are thus in a position to transfer some of the above conclusions for
V$_4$O$_7$ to V$_2$O$_3$.
First, V-V bonding across octahedral faces along $a_{\rm prut}$ seems to have the major
effect on the shape of the $d_{yz}$ DOS of V$_4$O$_7$ and thus on the shape
of the $a_{1g}$ DOS of V$_2$O$_3$. However, also the V-V hybridization across the faces parallel to
$b_{\rm prut}$ is large since it leads to strong splitting of the electronic states. Second, due
to the reduced metal-metal interaction across the octahedral edges parallel to $c_{\rm prut}$ the
vanadium $d_{x^2-y^2}$ states are substantially localized and undergo further localization at the transition.

This supports previous work by Dernier \cite{dernier70}, who concluded from a comparison of pure/doped V$_2$O$_3$
and of Cr$_2$O$_3$ that the metallic properties of V$_2$O$_3$ are connected to V-V
hybridization across shared octahedral edges rather than to hopping processes in the vanadium
pairs along $c_{\rm hex}$. From this point of view the MIT arises from structural changes occuring at the transition.
Via strong electron-phonon coupling the structural
distortions translate into a narrowing of the bands perpendicular to the $c_{\rm hex}$-axis. Note that the
decoupling of the vanadium atoms parallel to the $c_{\rm prut}$-direction is even stronger for the sesquioxide
than for V$_4$O$_7$ since the corresponding bond length shortens by $0.12$\,\AA\ instead of $0.06$\,\AA.
The band narrowing eventually paves the way for increased influence of electronic correlations
in the insulating configuration. So far our treatment of V$_2$O$_3$ concerned only the symmetry breaking PM-AFI
transition, which is accompanied by substantial modifications of the interatomic
V-V spacings. However, Pfalzer et al.\ \cite{pfalzer02} reported on local symmetry
breaking in the PI phase of aluminium doped V$_2$O$_3$, giving rise to similar structural and electronic
properties of the AFI and PI phase at least on a local scale. Due to the local distortions
one may expect a closely related scenario for both the PM-AFI and the PM-PI transition. In general, the reduction
of the dispersion perpendicular to $c_{\rm hex}$ seems to play a key role for the MIT of V$_2$O$_3$.

\section{Systematic aspects of the Magn\'eli series}
\label{sec3}
\begin{figure}[t!]\begin{center}
\includegraphics[width=10.0cm,clip]{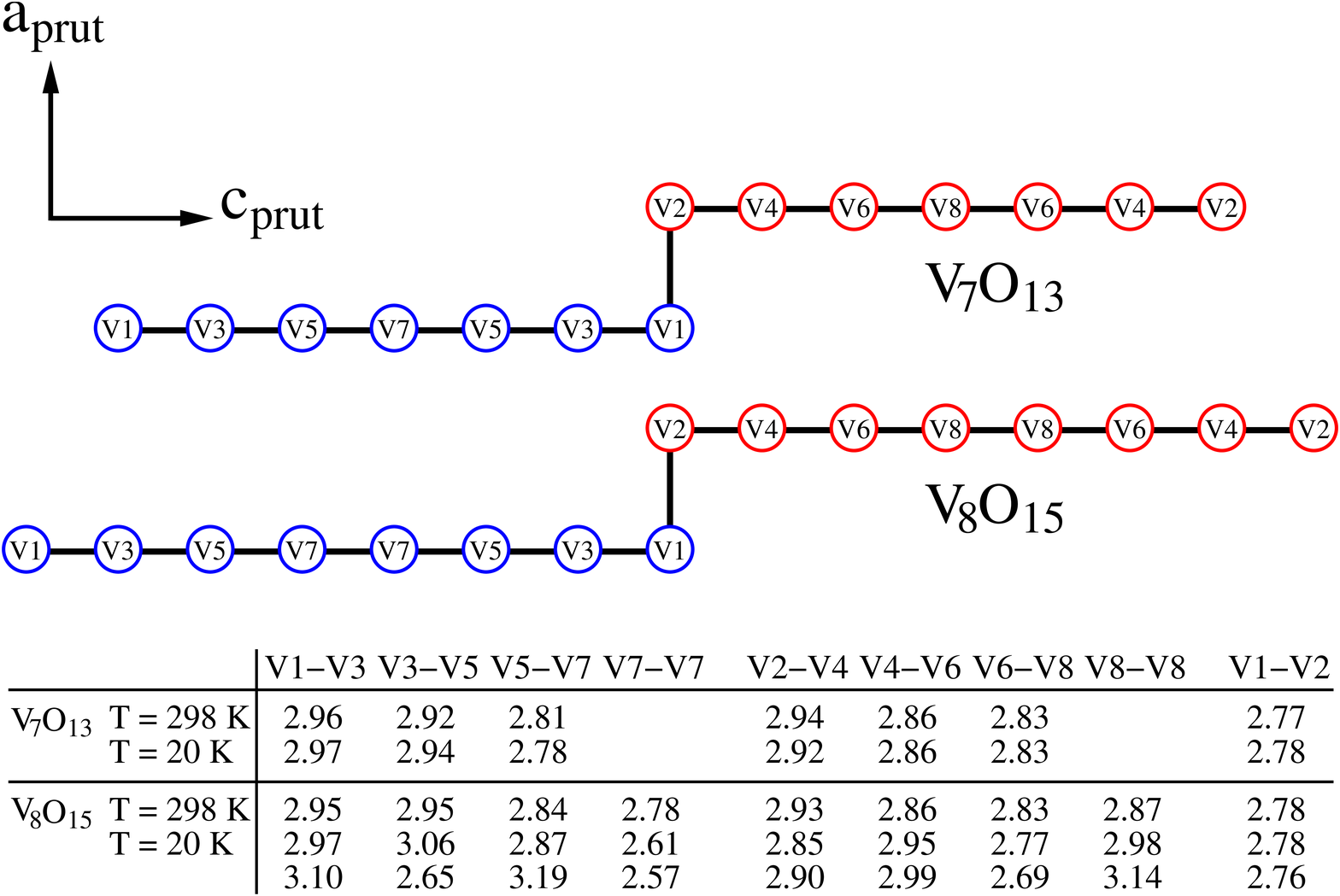}
\caption{Notation of the vanadium sites in V$_7$O$_{13}$ and V$_8$O$_{15}$, where
we observe metal chains of length $n=7$/$n=8$ along $c_{\rm prut}$.
In both cases eight inequivalent vanadium sites give rise to two types of chains. The spacial arrangement of the
chains is similar to that of V$_4$O$_7$ and V$_6$O$_{11}$, see figures
\ref{pic20} and \ref{pic19}. The table gives measured V-V distances (in \AA) for both the metallic (298\,K) and
the insulating (20\,K) phase of V$_7$O$_{13}$/V$_8$O$_{15}$ \cite{canfield90}. Below
the transition the latter compound develops a superstructure with four inequivalent chains.}
\label{pic35}\end{center}\end{figure}

Finally, addressing the systematics inherent in the Magn\'eli series we are interested in
effects of the chain length. Variation of the length systematically modifies the ratio of dioxide
and sesquioxide-like regions. In this context especially a comparison of V$_7$O$_{13}$ and V$_8$O$_{15}$
is useful because the materials are structurally very similar. Nevertheless, the former compound is
the only member of the Magn\'eli series which does not undergo an MIT as a function of temperature.
In a comparison of all Magn\'eli phases, VO$_2$ and V$_2$O$_3$ take an
exceptional position as they either miss chain end or chain center metal sites. In addition, they
display higher symmetry than that of the triclinic space group $P\bar{1}$ ($C_i^1$). This is likewise true for monoclinic
V$_3$O$_5$. The latter compound, although even closer related to the sesquioxide than V$_4$O$_7$, is therefore not
suitable for a detailed understanding of the electronic states in V$_2$O$_3$.
Both V$_5$O$_9$ and V$_8$O$_{15}$ are characterized by a superstructure doubling the
unit cell at the MIT. While the metal chains of low temperature V$_5$O$_9$
\cite{lepage91,marezio74} lose their inversion symmetry, there are four inequivalent chains for
V$_8$O$_{15}$. The notation of the vanadium sites in V$_7$O$_{13}$/V$_8$O$_{15}$ is given in figure \ref{pic35}.
For the former compound the chains are labeled V1-V3-V5-V7-V5-V3-V1 and V2-V4-V6-V8-V6-V4-V2, whereas for V$_8$O$_{15}$ the
innermost site is doubled, leading to the configurations V1-V3-V5-V7-V7-V5-V3-V1 and V2-V4-V6-V8-V8-V6-V4-V2.
Extra sites in the insulating superstructure are indicated by tildes. As in V$_4$O$_7$ and V$_6$O$_{11}$, the
oxygen sublattice of V$_8$O$_{15}$ hardly changes at the phase transition. The dominating structural modifications
are again concerned with the V-V dimerization along $c_{\rm prut}$, see the bond lengths in figure \ref{pic35}.

The study of V$_7$O$_{13}$/V$_8$O$_{15}$ is again based on LDA
calculations, which were performed analogous with the previously discussed Magn\'eli phases.
The partial DOS shows three groups of bands corresponding to O $2p$, V $3d$
$t_{2g}$, and V $3d$ $e_g^{\sigma}$ states. Due to a reduced formal electron count of $1.29$ and
$1.25$ electrons per metal site for the $n=7$ and $n=8$ compound, respectively, the filling of
the $t_{2g}$ group is slightly smaller than for V$_4$O$_7$ and V$_6$O$_{11}$.
Site-projected partial densities of states for V$_7$O$_{13}$ and V$_8$O$_{15}$ show the typical features discovered before.
In particular, the shape of the DOS curves is completely understood in terms of the vanadium
coordination, which is true for all Magn\'eli compounds. Hence it is useful to concentrate on
representative atomic sites: the central site V7 and the last but one site V3 of the 1-3-5-7 chain.

\begin{figure}[t!]\begin{center}
\includegraphics[width=10.0cm,clip]{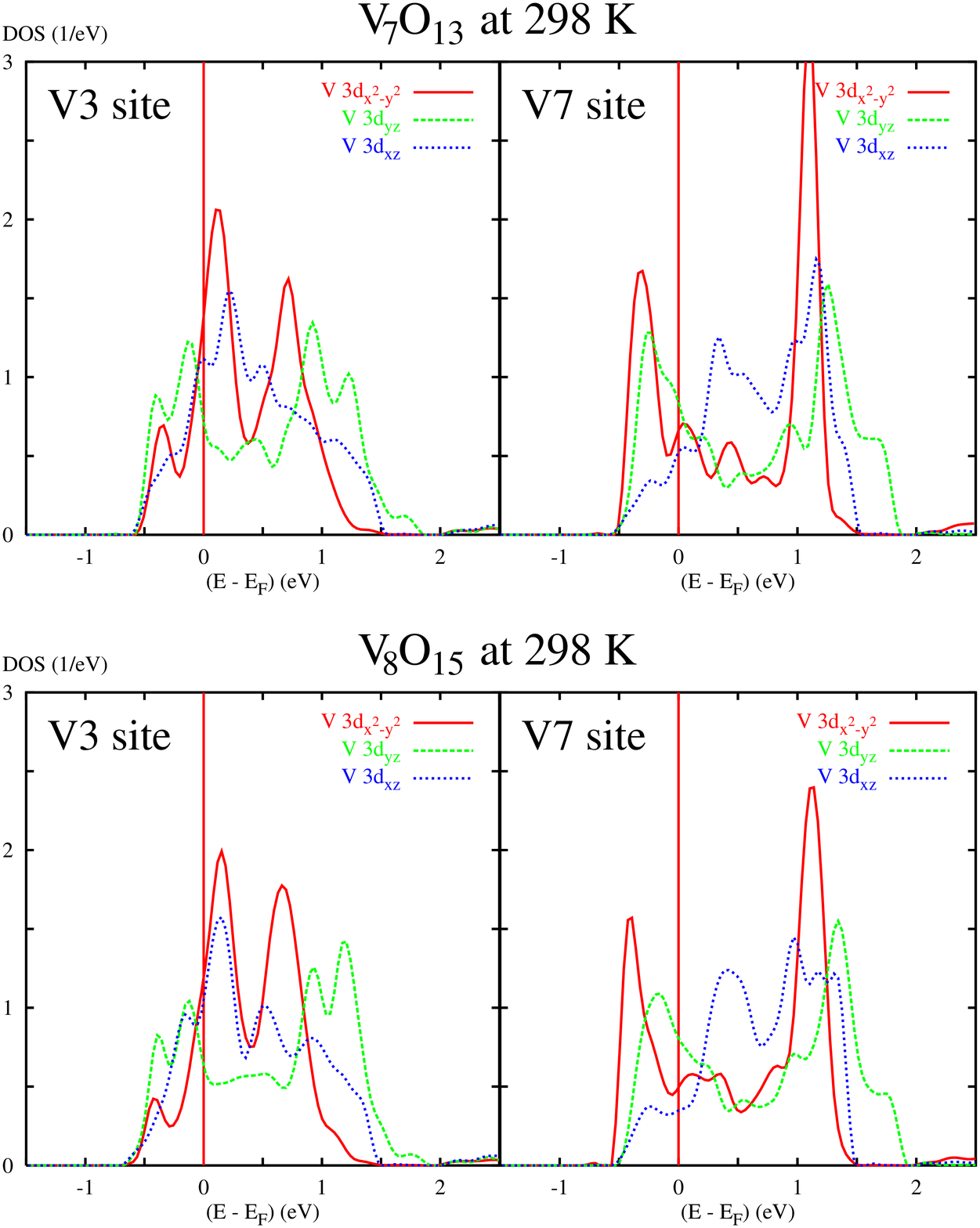}
\caption{Site-projected partial V $3d$ $t_{2g}$ densities of states (DOS) per vanadium atom for
high temperature V$_7$O$_{13}$ and V$_8$O$_{15}$. The orbitals refer to the local rotated reference
frame. Sites V3 and V7 reveal little discrepancy.}
\label{pic33}\end{center}\end{figure}

Figure \ref{pic33} shows site-projected partial densities of states for the sites V3 and V7
arising from the room temperature structures of V$_7$O$_{13}$ and V$_8$O$_{15}$.
The representation of the V $3d$ symmetry components refers to the local rotated reference frame with
the $z$ and $x$-axis parallel to the apical axis of the local octahedron and the $c_{\rm prut}$-axis,
respectively. V-O overlap combined with crystal field splitting results in two groups of antibonding states.
While the $t_{2g}$ dominated region reaches from about $-0.6$\,eV to $1.8$\,eV, one observes
additional $t_{2g}$ contributions above $2.0$\,eV, where the $e_g^{\sigma}$ states dominate. In consistence
with the positions of V$_7$O$_{13}$ and V$_8$O$_{15}$ within the Magn\'eli series, the latter are fairly small, which
resembles the behaviour of VO$_2$ rather than of V$_2$O$_3$. Contributions due to O $2p$ orbitals in the energy interval of
figure \ref{pic33}, which arise from covalent V-O bonding,  comprise less than
10\% of the total DOS at the Fermi level; they are not included in the figure. As a matter of fact,
the in-chain V-V distances in the high temperature modifications of V$_7$O$_{13}$ and V$_8$O$_{15}$
are very similar, see figure \ref{pic35}. Of course,
additional V7-V7 and V8-V8 bonds occur for chains of length $n=8$, which are not available for chains of length $n=7$.
The oxygen sublattice of the compounds is almost identical, which likewise applies to the
antiferroelectric-like lateral displacements of the metal atoms perpendicular to $c_{\rm prut}$ and the
rotation of the metal chains off the $c_{\rm prut}$-axis. Compared to V$_4$O$_7$ and V$_6$O$_{11}$ the latter
effects are weaker due to the dioxide-like nature of the long chain compounds. The importance
of the local atomic coordination for the DOS shape suggests a similar electronic structure
for both metallic V$_7$O$_{13}$ and V$_8$O$_{15}$, which is confirmed by figure \ref{pic33}.

The local $d_{x^2-y^2}$, $d_{yz}$, and $d_{xz}$ densities of states
resemble one another even with regard to the details of their shapes, which applies not only to the
sites V3 and V7 but also to all other metal atoms. Due to strong $\sigma$-type
metal-metal bonding along $c_{\rm prut}$ the chain center atoms are characterized by a distinct
splitting of the $d_{x^2-y^2}$ DOS. In the case of V7 we hence find sharp peaks at about
$-0.3$\,eV and $1.1$\,eV. Because of the increasing V-V distances at the chain ends the
bonding-antibonding splitting is noticeably smaller for the V3 $d_{x^2-y^2}$ DOS, where peaks occur
at about $0.2$\,eV and $0.7$\,eV. As a consequence of less efficient
$\pi$-type V-V overlap via the $d_{xz}$ orbitals the respective DOS is rather compact. In contrast,
a substantial splitting in bonding and antibonding branches can be found in the $d_{yz}$ DOS. For sites
V3 and V7 this shape traces back to metal-metal interaction along $b_{\rm prut}$.
The strength of the splitting is roughly identical for both sites. To conclude, the site-projected densities of
states of high temperature V$_7$O$_{13}$/V$_8$O$_{15}$ are well understood from
the discussions in the preceding sections. By virtue of very closely related metal-metal bond lengths
in the compounds similar DOS curves arise. Thus it is
\begin{figure}[t!]\begin{center}
\includegraphics[width=10.0cm,clip]{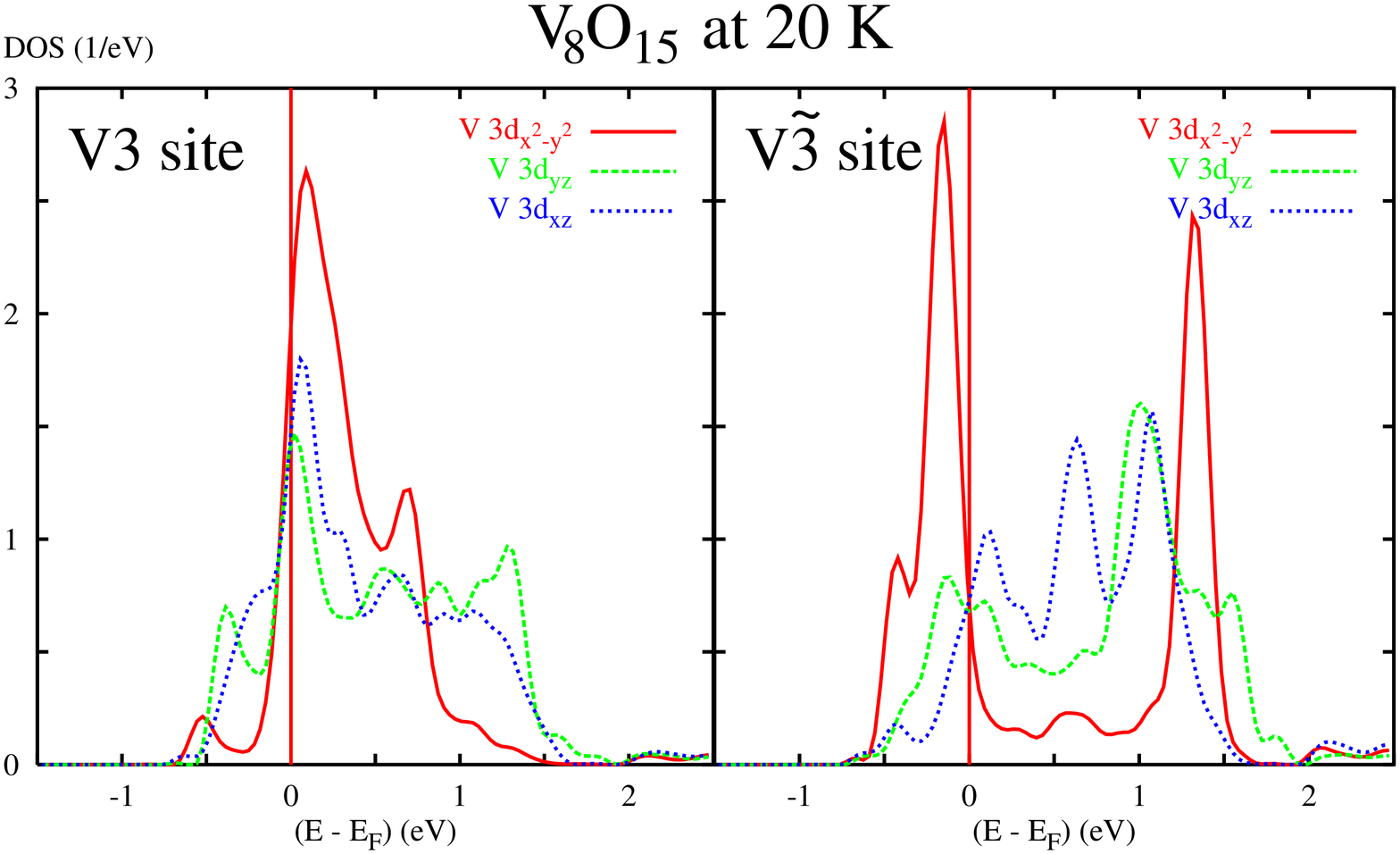}
\caption{Site-projected partial V $3d$ $t_{2g}$ densities of states (DOS) per metal atom for
sites V3 and V${\tilde{3}}$ of low temperature V$_8$O$_{15}$. At high temperatures these sites
are equivalent. The orbitals refer to the local rotated reference frame.}
\label{pic34}\end{center}\end{figure}
surprising that only one of the materials undergoes an MIT.
While the charges in the 1-3-5-7 chain of V$_7$O$_{13}$ and the 2-4-6-8 chain of V$_8$O$_{15}$ are more
or less equally distributed among the metal sites, the other chains show a small shift
of charge towards the chain centers, but generally such charge transfer effects are very small.
In the calculation the metallic phases of V$_7$O$_{13}$ and V$_8$O$_{15}$ are
barely distinguishable as far as the local electronic behaviour is concerned.

The investigation of the low temperature V$_8$O$_{15}$ modification reveals the same mechanisms
found previously in the Magn\'eli series. Depending on the bond lengths to the adjacent vanadium sites
along $c_{\rm prut}$ one finds either increased splitting or enhanced localization of the $d_{x^2-y^2}$
orbitals. The V-V bond lengths indicate strong V-V dimerization in the
$\tilde{1}$-$\tilde{3}$-$\tilde{5}$-$\tilde{7}$, 2-4-6-8, and $\tilde{2}$-$\tilde{4}$-$\tilde{6}$-$\tilde{8}$
chain, but not in the 1-3-5-7 chain. While the innermost V-V distances of both the 2-4-6-8 and
$\tilde{2}$-$\tilde{4}$-$\tilde{6}$-$\tilde{8}$ chain
elongate at the MIT from $2.87$\,\AA\ to $2.98$\,\AA\ and $3.14$\,\AA, they shorten for the other chains
from $2.78$\,\AA\ to $2.61$\,\AA\ and $2.57$\,\AA. Thus the dimerization patterns are
inversed. As a consequence, in the low temperature structure all vanadium sites, except for V1, V3, V5,
and V$\tilde{1}$, are involved in strong V-V overlap along $c_{\rm prut}$. The increased
interaction leads to a stronger bonding-antibonding splitting of the $d_{x^2-y^2}$ DOS.
For atoms affected by V-V dimerization one additionally obtains energetical upshifts of the
$d_{yz}$ and $d_{xz}$ densities of states. In accordance with earlier findings they
show typical changes in the electronic structure known from the MIT of VO$_2$. Although
the energetical separation between the $d_{x^2-y^2}$ states and the remaining $e_g^{\pi}$ states of
V$_8$O$_{15}$ grows in the low temperature phase it is still far from complete.
The described Peierls-like behaviour does not
account for sites V1, V3, V5, and V$\tilde{1}$, which are characterized by decreasing V-V overlap
below the MIT. Due to the reduced metal-metal bonding the $d_{x^2-y^2}$ densities of states
sharpen and develop a pronounced peak close to the Fermi level. Because of their
sesquioxide-like coordination the behaviour is not surprising for sites V1 and V$\tilde{1}$. However, the
separation from neighbouring vanadium atoms forces V3 and V5 to display the same localization of the
$d_{x^2-y^2}$ orbitals. Hence electronic correlations may be as important for these states as for the
V$_2$O$_3$-type chain end sites. The question why one of the four vanadium chains in the low temperature
V$_8$O$_{15}$ superstructure prefers localization to dimerization is difficult to answer. The calculated V
$3d$ valence charges do not give rise to any systematics. Partial
$t_{2g}$ densities of states are depicted in figure \ref{pic34}. For the dioxide-like V$\tilde{3}$
site we recognize both an increased splitting of the $d_{x^2-y^2}$ DOS and an energetical upshift
of the $d_{yz}$ and $d_{xz}$ curves. In contrast, the structurally separated V3 site is characterized by
a very compact $d_{x^2-y^2}$ DOS located almost completely above the Fermi level. In comparison
to the sesquioxide-like V1 site of V$_4$O$_7$, see figure \ref{pic17}, the filling of the low temperature
$d_{x^2-y^2}$ states is significantly smaller due to the reduced electron count. Summarizing, these
findings support our MIT scenario as a mixture
of features typical of either vanadium dioxide or sesquioxide, which we developed in the previous sections.

We are nevertheless left with the question why the compound V$_7$O$_{13}$ does not undergo an MIT. From purely structural considerations
the main difference between V$_7$O$_{13}$ and V$_8$O$_{15}$ is the amount of dioxide-like center sites.
The local environments and thus the electronic properties of the individual sites are very similar.
However, we point out that the MITs in the Magn\'eli series arise as combinations of an embedded Peierls instabilty
inherent in the metal chains and electronic correlations relevant for the
chain end atoms. Since compounds with small vanadium chains are dominated by chain
end sites and thus more susceptible to correlations, their MITs are most likely driven by sesquioxide-like states.
On the other hand, the Peierls mechanism should be dominating for materials with long chains.
This scenario of different driving forces of the MITs is consistent with the behaviour of the transition temperatures
in the Magn\'eli series as seen in figure \ref{pic36}. With only one exception ($n=5$) the
temperatures decrease monotonously from V$_3$O$_5$ to V$_7$O$_{13}$, which does not show a transition.
In the case of V$_5$O$_9$ the peculiar superstructure evolving below the MIT might cause the observed
deviation. Transferred to our picture this kind of systematics corresponds to the decreasing influence of chain
end sites as compared to the growing impact of the chain centers. When the chain length increases the influence of the local
electronic correlations at the chain ends diminishes. At the same time the Peierls-like mechanism becomes stronger. 
With an increasing number of chain center atoms available for V-V dimerization the energy gain resulting from such a
modification becomes more and more advantageous. With the experimental transition temperatures in mind we thus
attribute the MIT of V$_8$O$_{15}$ to a Peierls-like scheme; for V$_7$O$_{13}$ neither
of the two mechanisms is strong enough to induce a transition.

\section{Conclusions}
The vanadium Magn\'eli phases V$_n$O$_{2n-1}$ form a homologous series of compounds with crystal
structures comprising dioxide and sesquioxide-like regions. Studying the MITs in this material
class thus paves the way for a more complete understanding of the phase transitions of the prototypical oxides
VO$_2$ ($n\to\infty$) and V$_2$O$_3$ ($n=2$). Analyzing those V $3d$ $t_{2g}$
states which are involved in the phase transitions allows us to discuss the changing influence of electronic correlations
and electron-lattice interaction on going from VO$_2$ to V$_2$O$_3$. We have developed
a unifying description of the various crystal
structures including both the rutile structure of VO$_2$ and the corundum structure of V$_2$O$_3$. Our
systematic representation is based on a regular three-dimensional network of oxygen octahedra partially
filled by metal atoms, giving rise to characteristic metal chains of length $n$.
Due to the comprehensive picture of the structures, the
electronic bands can be grouped into states behaving either VO$_2$ or V$_2$O$_3$-like.
The detailed electronic structure and consequently the MITs of the
Magn\'eli phases turn out to be strongly influenced by the local metal-metal coordination.
The phase transitions arise as results of both electron lattice interaction within the dioxide-like and
electronic correlations within the sesquioxide-like regions of the crystal. Dioxide-like vanadium sites
reveal the characteristic features of the embedded Peierls instability responsible for the
MIT of VO$_2$. The combination of dimerization and antiferroelectric-like displacements of the metal
atoms by means of strong electron-lattice interaction causes splitting of the $d_{x^2-y^2}=d_{\parallel}$
states and energetical upshift of the $d_{yz}$/$d_{xz}$ states. Sites related to V$_2$O$_3$ are
characterized by strongly reduced V-V overlap. Thus the $d_{x^2-y^2}$ states become more localized.
Consequently, correlation effects play a more important role for these states. In conclusion, the electronic structures
and the metal-insulator transitions of
the Magn\'eli phases remain a crucial test case for theories aiming at a correct description of both VO$_2$ and V$_2$O$_3$.

\begin{acknowledgement}
We are grateful to U.~Eckern for many fruitful discussions as well as for his continuous support. Discussions with
S.~Horn, S.~Klimm, T.~Kopp, P.~Pfalzer, and K.~Schwarz are acknowledged.
This work was supported by the Deutsche Forschungsgemeinschaft (Sonderforschungsbereich 484, Augsburg).
\end{acknowledgement}

\end{document}